\documentclass[12pt]{article}
\usepackage{authblk}
\usepackage{graphics}
\usepackage{graphicx}
\usepackage{amssymb}
\usepackage{amsmath}
\usepackage{adjustbox}
\usepackage{cite}
\usepackage[usenames]{color}
\usepackage[labelfont={bf,sf}]{caption}
\usepackage[caption=false,labelformat=simple]{subfig}

\def\abstract{\topsep=0pt\partopsep=0pt\parsep=0pt\itemsep=0pt\relax
	\trivlist\item[\hskip\labelsep
	{\bfseries\abstractname}]\if!\abstractname!\hskip-\labelsep\fi
}
\def\abstractname{Abstract.}

\usepackage{titling}

\thanksmarkseries{alph}

\makeatletter\renewcommand{\@biblabel}[1]{#1.}\makeatother

\usepackage{lipsum}

\let\OLDthebibliography\thebibliography
\renewcommand\thebibliography[1]{
	\OLDthebibliography{#1}
	\setlength{\parskip}{0pt}
	\setlength{\itemsep}{0pt plus 0.3ex}
}

\def\figurename{\textbf{Fig}.}
\def\fnum@figure{\figurename\thinspace\thefigure}
\makeatletter
\def\email#1{{\tt#1}}
\def\keywordname{\textbf{Keywords:}}
\providecommand{\keywords}[1]{\par\addvspace\baselineskip
	\noindent\keywordname\enspace\ignorespaces#1}
\def\bbbc{{\mathchoice {\setbox0=\hbox{$\displaystyle\rm C$}\hbox{\hbox
				to0pt{\kern0.4\wd0\vrule height0.9\ht0\hss}\box0}}
		{\setbox0=\hbox{$\textstyle\rm C$}\hbox{\hbox
				to0pt{\kern0.4\wd0\vrule height0.9\ht0\hss}\box0}}
		{\setbox0=\hbox{$\scriptstyle\rm C$}\hbox{\hbox
				to0pt{\kern0.4\wd0\vrule height0.9\ht0\hss}\box0}}
		{\setbox0=\hbox{$\scriptscriptstyle\rm C$}\hbox{\hbox
				to0pt{\kern0.4\wd0\vrule height0.9\ht0\hss}\box0}}}}
			
\definecolor{lightsg}{rgb}{0.13, 0.7, 0.67}
\definecolor{berry}{rgb}{0.89, 0.04, 0.36}

\title{\textsf{Spectral manipulation of the trigonometric Rosen-Morse potential through supersymmetry}}

\author{{\normalsize David J. Fern\'andez} \thanks{e-mail:\email {david@fis.cinvestav.mx}} \enspace {\normalsize and Rosa Reyes}
	\thanks{e-mail: \email {rmreyes@fis.cinvestav.mx} (corresponding author)}}

\affil{\footnotesize Physics Department, Cinvestav, AP 14-740, Mexico City 07000, Mexico}

\date{}
\begin{document}
{\setlength{\parskip}{-2cm}
\maketitle}
\thispagestyle{empty}

\abstract{	
The first and second-order supersymmetry transformations can be used to manipulate one or two energy levels of the initial spectrum when generating new exactly solvable Hamiltonians from a given initial potential. In this paper, we will construct the first and second-order supersymmetric partners of the trigonometric Rosen-Morse potential. Firstly, it is identified a set of solutions of the initial stationary Schr\"odinger equation which are appropriate for implementing in a simple way non-singular transformations, without inducing new singularities in the built potential. Then, the way the spectral manipulation works is illustrated through several specific examples.
}
\section{Introduction}
\label{intro}
Exactly solvable quantum potentials has deserved physicist attention for many years \cite{mi68,lr91}. In order to generate this kind of potentials, supersymmetric quantum mechanics (SUSY QM) has been recurrently  employed \cite{CKS95,Jun96,Bag00,Fer05,Fer10,Fer19}. This approach is equivalent to the factorization method, intertwining technique and Darboux transformation \cite{abi84,bs97}. After applying the procedure, it turns out that the spectra of the initial and final Hamiltonians can differ in a finite number of levels, thus suggesting the possibility to implement in this way the so-called spectral design \cite{zc97,fr01,gns03,ac04,fr20}.

On the other hand, the short range potentials are specially important in high energy physics, since they could be used to model quark confinement in quantum chromodynamics (QCD). The trigonometric Rosen-Morse (TRM) potential belongs to this class \cite{Ros32,MS19}, \textit{e.g.}, in QCD it has been proposed for describing the quark-gluon interaction \cite{Kir07}, and also as an effective potential for quark-antiquark interaction \cite{Abu19}. The TRM potential has been employed as well to describe thermodynamic properties of a quantum gas of mesons, because it provides a reasonable parametrization of the mesons internal structure \cite{Bah20}. The TRM potential arises also in the study of exact solutions for the position-dependent mass Schr\"odinger equation \cite{Ram03}, and as a proposal for studying rotational type degeneration in diatomic molecules described by means of a cotangent-perturbed rigid rotator \cite{ack11}.

It could be important as well in mathematical physics, in the study of non-linear algebras and SUSY QM \cite{fhr07,ChF08,Zel18}, since there is a non-linear relation between the energy eigenvalues and the quantum number characterizing them. In addition, the TRM potential appears quite naturally when addressing quantum potentials on curved spaces \cite{nsr99}. Let us mention also the use of the TRM potential in studies of graphene in external magnetic fields, where it appears when considering the trigonometric singular well $\mathbf{B}=(0,0,B_0/\sin^2 (x))$ \cite{knn09,fgo20,fgo21}. The SUSY transformations have been recently applied to the TRM potential in order to identify the system ladder operators and construct coherent states \cite{Gar21}.

Although the SUSY techniques have been implemented for plenty of exactly solvable potentials \cite{Fer10,fhr07,fgn98,fhm98,fh99,cf08}, up to our knowledge the method has not been applied thoroughly to the trigonometric Rosen-Morse potential. In fact, the first-order SUSY has been used to delete the ground state energy level, but the shape-invariance of the TRM potential makes the new potential to be essentially the same as the initial one \cite{Bar91,CKS95,ck06}. The second-order SUSY has been as well implemented, by deleting two consecutive energy levels of the TRM potential \cite{Dom11}. However, in both cases the employed seed solutions are bound state eigenfunctions of the TRM Hamiltonian. It would be important to determine if the SUSY method can be applied through general seed solutions, which are not necessarily bound state eigenfunctions of the initial Hamiltonian. 

In this paper, we are going to study the first and second-order SUSY partners of the tri\-go\-no\-me\-tric Rosen-Morse potential. These transformations will allow us to design the spectra for the new potentials in several different ways, as will be shown throughout this article. The seed solutions chosen to implement the transformations will be more general than in previous works \cite{CKS95,ck06,Dom11}, for factorization energies which are not necessarily associated to bound states.

This paper has been organized as follows. In Sect.~\ref{SUSY} we will make a brief overview of SUSY QM. In Sect.~\ref{TRM} we will discuss the TRM potential, the construction of the appropriate solutions of the corresponding Schr\"odinger equation and the analysis of them. In Sect.~\ref{SUSY partners} we will show the results of applying the SUSY techniques to the TRM potential. Finally, our conclusions will be presented in Sect.~\ref{Conclusions}.

\section{Supersymmetric quantum mechanics}
\label{SUSY}

Supersymmetric quantum mechanics is a powerful tool to generate exactly solvable potentials from a given initial one. The spectrum of the new potential depends on the order of the differential operator used to implement the transformation. Since in this paper we will consider first and second-order transformations, we will give now a brief overview of both SUSY cases.

\subsection{First-order SUSY QM}
\label{1susyqm}

Let us consider the two intertwining relationships
\begin{eqnarray}
&& H_1 A_1^+ = A_1^+ H_0, \\
&& H_0 A_1^- = A_1^- H_1 ,
\end{eqnarray}
where $A_1^\pm$ are the first-order differential operators
\begin{eqnarray}
&& A_1^\pm = \frac{1}{\sqrt{2}}\left(\mp\frac{d}{dx} + \frac{u_{01}'(x)}{u_{01}(x)}\right), 
\end{eqnarray}
and $H_0, H_1$ are two Schr\"odinger Hamiltonians,
\begin{eqnarray}
&& H_i = - \frac12\frac{d^2}{dx^2} + V_i(x), \quad i=0,1. \label{SchrHam}
\end{eqnarray}
We suppose that the initial potential $V_0(x)$ is defined in the interval $x\in(x_L,x_R)$, which includes the possibility that this domain could be the real line, the positive real line or some finite interval.
The SUSY partner potentials $V_0(x), \ V_1(x)$ and the {\it seed solution} $u_{01}(x)$ turn out to be related 
\cite{CKS95,Jun96,Bag00,Fer05,Fer10,Fer19}:
\begin{eqnarray}
&& V_1(x) = V_0(x) - [\ln u_{01}]'' , \label{npss} \\
&& H_0 u_{01} = - \frac12 {u_{01}}''  + V_0 u_{01} = \epsilon_1 u_{01}. \label{se}
\end{eqnarray}
The formal eigenfunction $u_{01}$ of $H_0$ associated to the {\it factorization energy} $\epsilon_1$ satisfies as well that $A_1^+ u_{01} = 0$. The previous equations imply the factorization of $H_0$ and $H_1$ as follows:
\begin{eqnarray}
&& H_0 = A_1^- A_1^+ + \epsilon_1 , \\
&& H_1 = A_1^+ A_1^-  + \epsilon_1 .
\end{eqnarray}

Suppose that $H_0$ is an exactly solvable Hamiltonian, with normalized eigenfunctions $\psi_{0n}(x)$ and eigenvalues $E_n$, $n = 0,1,\dots$ which are given. In order to implement the first-order SUSY, we will use a nodeless seed solution $u_{01}(x)$ associated with a factorization energy $\epsilon_1$ such that $\epsilon_1 \leq E_0$, so that none new singularity will appear in $V_1(x)$ (see Eq.(\ref{npss})). Thus, the normalized eigenfunctions $\psi_{1n}(x)$ of $H_1$ associated to the eigenvalues $E_n$ acquire the form:
\begin{equation}
\psi_{1n}(x) = \frac{A_1^+ \psi_{0n}(x)}{\sqrt{E_n-\epsilon_1}}.
\end{equation}
There is also a formal eigenfunction $\psi_{1\epsilon_1}$ of $H_1$ associated to $\epsilon_1$ which is annihilated by $A_1^-$, \textit{i.e.}, $H_1 \psi_{1\epsilon_1}(x) = \epsilon_1 \psi_{1\epsilon_1}(x)$, 
$A_1^- \psi_{1\epsilon_1}(x) = 0$. Thus,
\begin{eqnarray}
&& \psi_{1\epsilon_1}(x) \propto \frac{1}{u_{01}(x)}. 
\end{eqnarray}
Since the spectrum of $H_1$ will depend on the square-integrability of $\psi_{1\epsilon_1}(x)$, three different cases appear.

\smallskip

\noindent i) Deleting a level. If $\epsilon_1 = E_0$ and $u_{01}(x) = \psi_{00}(x)$, it turns out that $\psi_{1\epsilon_1}(x)$ is not square-integrable, thus $E_0 \notin {\rm Sp}(H_1)$ and hence
\begin{eqnarray}
& {\rm Sp}(H_1) = \{E_n, n=1,2,\dots\} = {\rm Sp}(H_0)-\{E_0\}.
\end{eqnarray}

\noindent ii) Creating a new level. If $\epsilon_1 < E_0$ with an associated nodeless seed solution $u_{01}(x)$, it turns out that $\psi_{1\epsilon_1}(x)$ is square-integrable and then $\epsilon_1\in{\rm Sp}(H_1)$, \textit{i.e.},
\begin{eqnarray}
& {\rm Sp}(H_1) = \{\epsilon_1, E_n, n=0,1,\dots\} = \{\epsilon_1\}\cup{\rm Sp}(H_0).
\end{eqnarray}
Note that $\psi_{1\epsilon_1}(x)$ is called sometimes `missing state' \cite{mi84} (see also \cite{fgn98}).

\noindent iii) Isospectral transformation. If $\epsilon_1 < E_0$ but the seed solution $u_{01}(x)$ has a node at one of the ends of the potential domain (if the end is infinite this node is achieved as a limit process), it turns out that $\psi_{1\epsilon_1}(x)$ is not square-integrable and then $\epsilon_1\notin{\rm Sp}(H_1)$, \textit{i.e.},
\begin{eqnarray}
& {\rm Sp}(H_1) = \{E_n, n=0,1,\dots\} = {\rm Sp}(H_0).
\end{eqnarray}

\subsection{Second-order SUSY QM}
\label{2susyqm}

The second-order SUSY QM is based on the intertwining relationship
\cite{ais93,aicd95,fe97,sa99,pl00,ast01,pl04}
\begin{equation}
H_2 B_2^{+}=B_2^{+} H_0,
\label{IntertwineRelation}
\end{equation}
where $B_2^{+}$ is the second-order differential operator 
\begin{equation} 
B_2^{+}=\frac{1} {2} \left(\frac {\mbox{d}^2}{\mbox{d}x^2}-\eta (x)\frac {\mbox{d}}{\mbox{d}x}+\gamma (x)\right), 
\end{equation}
with $\eta (x)$ and $\gamma (x)$ being two real functions to be found. The other operators of Eq.~(\ref{IntertwineRelation}) are two Schr\"odinger Hamiltonians $H_i, i=0,2$ whose form is given in Eq.~(\ref{SchrHam}). Note that, as for the first-order SUSY QM, there is a complementary intertwining relationship
\begin{equation}
H_0 B_2^{-}=B_2^{-} H_2,
\label{IntertwineRelation2}
\end{equation} 
where $B_2^{-}= (B_2^{+})^\dagger$, which arises from calculating the Hermitian conjugate of Eq.~(\ref{IntertwineRelation}). If $B_2^{+}$ and $H_i, \, i=0,2$ are replaced in Eq.~(\ref{IntertwineRelation}), after some work the following system of equations appears \cite{Fer05}:
\begin{eqnarray}
&& V_2=V_0-\eta' , \\ 
&& \gamma=\frac{\eta'}2+\frac{\eta^2}2-2V_0+d, \\
\label{system1}
&& \frac {\eta \eta''} 2 -\frac{ \eta'^{2}}4 + \eta^2 \eta' + \frac{\eta^{4}}4 - 2V_0\eta^2 + d\eta^2 + c = 0,
\label{etaequation}
\end{eqnarray}
where $d$, $c$ are two real integration constants. In order to solve this system of equations more information is required, since we have only three equations to determine four functions, $V_0, V_2, \ \eta, \ \gamma$. 

As can be seen, if $V_0(x)$ is given it is possible to determine $V_2(x)$ and $\gamma (x)$, once the solution $\eta(x)$ to the non-linear differential Eq.~(\ref{etaequation}) is obtained. This is achieved through the following \textit{ansatz} (see \cite{Fer05})
\begin{equation*}
\eta'=- \eta^2 +2\beta \eta +2\xi,
\end{equation*}
with $\beta$, $\xi$ being two functions of $x$ to be  determined. This ansatz transforms Eq.~(\ref{etaequation}) into the following set of equations 
\begin{eqnarray*}
	& \xi^2 =c,\qquad \epsilon=\frac 12 (d+\xi),\\ 
	& \beta'+ \beta^2 =2\left(V_0-\epsilon \right).
\end{eqnarray*}
The first two equations produce $ \xi_{1,2} =\pm \sqrt c$ and $\epsilon_{1,2} =\left( d\pm \sqrt c\right)/2$. The third one is a first-order non-linear differential equation, known as Riccati equation, which can be transformed into a linear one through the change $\beta_i=u'_{0i}/u_{0i}$, leading to
\begin{equation}
H_0 u_{0i} = -\frac{1} 2 u_{0i}''+V_0 u_{0i}=\epsilon_i u_{0i}, \qquad i=1,2.
\label{Schrod1}
\end{equation}
This is the stationary Schr\"odinger equation for the potential $V_0$ with factorization energies $\epsilon_1, \ \epsilon_2$. The functions $u_{0i}$, $i=1,2$ are called either transformation functions or seed solutions in the literature \cite{Fer05,Fer10,Fer19,Dom11}, they satisfy the Schr\"odinger equation~(\ref{Schrod1}) and belong to the kernel of $B_2^{+}$, \textit{i.e.}, \mbox{$B_2^{+}u_{0i}=0$}, $i=1,2$. Depending on whether they are square~integrable or not, they will be called physical or non-physical solutions of the initial Hamiltonial $H_0$.

The transformations of our interest in this paper are non-singular, \textit{i.e.}, the new potentials cannot have new singularities inside the domain of definition of the initial potential. The conditions to fulfill this requirement have been widely studied by years (see \textit{e.g.} \cite{CKS95,Jun96,Bag00,Fer05,Fer10,Fer19}). Moreover, the non-singular transformations have been classified according to the sign of the constant $c$ defining the factorization energies $\epsilon_{1,2}$. Consequently, three different cases appear: the real case $\epsilon_i\in{\rm I\!R}, \, i=1,2$ for $c>0$; the complex case $\epsilon_i\in\setbox0=\hbox{\rm C}\hbox{\hbox
	to0pt{\kern0.4\wd0\vrule height0.9\ht0}\box0}, \, i=1,2$ for $c<0$; the confluent case $\epsilon_1=\epsilon_2=d/2\in{\rm I\!R}$ for $c=0$.  Each case leads to specific restrictions on the seed solutions chosen to implement the transformation, according to the kind of modified spectrum that the new Hamiltonian will have with respect to the initial one.

\subsubsection{Real case}

If $c>0$ the two factorization energies $\epsilon_1, \, \epsilon_2$ turn out to be real and different; we have ordered them as $\epsilon_1>\epsilon_2$. The resulting potential reads:
\begin{equation}\label{newVreal}
V_2(x)=V_0(x)- \left[\ln\left(W(u_{01},u_{02})\right)\right]'',
\end{equation}
where $W(u_{01},u_{02})$ is the Wronskian of the two seed solutions $u_{01}$ and $u_{02}$. In order to avoid added singularities in the new potential $V_2(x)$, this Wronskian should be nodeless. If both $\epsilon_1,\,\epsilon_2$ do not coincide with bound state energies, this condition is fulfilled if $u_{02}$ has an extra node with respect to $u_{01}$ \cite{Fer05}. In particular, if both factorization energies are below the ground state energy $E_0$, then $u_{01}$ should be nodeless and $u_{02}$ should have one node, while for $\epsilon_1, \, \epsilon_2$ above $E_0$ both should fall in the same energy gap $(E_j,E_{j+1}]$ with the associated seed solutions $u_{02}$ and $u_{01}$ having necessarily $j+2$ and $j+1$ nodes respectively. On the other hand, if the two factorization energies belong to the initial spectrum, the Wronskian $W(u_{01},u_{02})$ will be nodeless only for $\epsilon_1=E_{j+1}$ and $\epsilon_2=E_{j}$.

The eigenfunctions $\psi_{0n}(x)$ of the initial Hamiltonian $H_0$ are related with those $\psi_{2n}(x)$ of the new Hamiltonian $H_2$ as follows: 
\begin{equation}\psi_{2n}\left(x\right)= \frac{B_2^{+} \psi_{0n}\left(x\right)}{\sqrt{\left(E_n-\epsilon_1\right)\left(E_n-\epsilon_2\right)}}.
\label{newPFreal}
\end{equation}
Moreover, there are explicit solutions to the stationary Schr\"odinger equation for $H_2$ with factorization energies $\epsilon_1, \, \epsilon_2$, which are given by
\begin{equation}
\quad \psi_{2\epsilon_1} \left(x\right)\propto \frac{u_{02}\left(x\right)} {W(u_{01},u_{02})},\qquad \psi_{2\epsilon_2} \left(x\right)\propto \frac{u_{01}\left(x\right)} {W(u_{01},u_{02})}.
\label{newEpFreal}	
\end{equation}
%

It is clear now that the kind of spectral modifications that could appear for the resultant Hamiltonian $H_2$ compared with the initial spectrum, depends on the factorization energies $\epsilon_1, \, \epsilon_2$ chosen, as well as on the square-integrability of $\psi_{2\epsilon_1}$ and $\psi_{2\epsilon_2}$.

\subsubsection{Complex case}

For $c<0$ the two factorization energies $\epsilon_1, \, \epsilon_2$ become complex conjugate to each other, $\epsilon_2=\epsilon^*_1$, which is the reason of the name {\it complex case} \cite{fmr03}. The new potential is given by 
\begin{equation}
V_2(x)=V_0(x)- \left[\ln\left(W(u_{01},u^*_{01})\right)\right]'',
\end{equation}
where $u_{01}$ and $u^*_{01}$ are two complex conjugate seed solutions of Eq.~(\ref{Schrod1}). 
The condition for producing non-singular transformations is that $u_{01}$ fulfills (see for example \cite{Fer05,Fer10,Fer19})
\begin{equation}
\lim_{x \to x_L}u_{01}(x)=0 \qquad \text{or}\qquad\lim_{x \to x_R}u_{01}(x)=0 .
\label{Limitcomplex}
\end{equation}
These transformations turn out to be isospectral.

\subsubsection{Confluent case}

The confluent case appears for $c=0$, when we get just a single factorization energy $\epsilon_1=\epsilon_2\in{\rm I\!R}$. In this case the new potential is once again given by Eq.~(\ref{newVreal}), but now
\begin{equation}
W(u_{01},u_{02})=w_0+\int_{x_0}^x{\left[u_{01}(y)\right]^2\mbox{d}y} ,
\label{confluent}
\end{equation} 
where $u_{01}$ satisfies the Schr\"odinger equation~(\ref{Schrod1}) but $u_{02}$ fulfills now $(H_0-\epsilon_1)u_{02} = u_{01}$, and $w_0$ is an integration constant that can be adjusted to avoid that $W(u_{01},u_{02})$ has a node in the $x$-domain of $V_0(x)$. Similarly as in the complex case, the seed solution $u_{01}$ must fulfill the conditions of Eq.~(\ref{Limitcomplex}) (see for example \cite{Fer05,Fer10,Fer19,fr20,fs03,fs05,fs11,bff12,cjp15,gq15,cs15,be16,cs17}).

The kind of spectral changes that can be achieved for $H_2$ depends on the chosen factorization energy $\epsilon_1$, as well as on the value of $w_0$. Hence, it is possible to generate a new Hamiltonian $H_2$ which is isospectral to $H_0$, with an additional energy level or without a given initial one as compared with the initial spectrum.

\section{Trigonometric Rosen-Morse potential}
\label{TRM}

The trigonometric Rosen-Morse potential is given by 
%
\begin{equation}
V(x) =\frac{a(a+1)} 2 \csc^2(x) -b \cot (x), \qquad a>0,\qquad b\in {\rm 
	I\!R,}
\label{tRMp}
\end{equation}
which is defined on the finite interval $x\in(0,\pi)$, i.e., $x_L=0$ and $x_R=\pi$. We need to solve first the stationary Schr\"odinger equation $H\psi(x)=E\psi(x)$, which in dimensionless coordinates reads
\begin{equation}
\left(-\frac{1} 2 \frac {\mbox{d}^2}{\mbox{d}x^2}+\frac{a(a+1)} 2 \csc^2(x) -b \cot (x)\right) \psi (x)=E\psi (x).
\label{StacionaryEq}
\end{equation}
 This equation, as well as the one for the original Rosen-Morse potential, can be transformed into the hypergeometric equation \cite{Ros32,Gan94}. For doing that, let us express $\psi(x)$ as follows:
\begin{equation}
\psi(x) =e^{-\frac {\mu} 2 x}\left(1+\cot^2(x)\right)^{(\nu-1)/2}f(x).
\label{Factorization1}
\end{equation}
%
If in the equation that results for $f(x)$ it is made the change of variable $z=\cot(x)$ we arrive at
\begin{equation}
(1+z^2)\frac {\mbox{d}^2 f}{\mbox{d}z^2} + (\mu + 2\nu z) \frac {\mbox{d} f}{\mbox{d}z} + [\nu(\nu-1) - a(a+1)]f=0 ,
\label{intermedia}
\end{equation}
where we have supposed that
\begin{equation}
\mu(\nu - 1)+ 2 b = 0, \qquad \frac{\mu^2}4 - (\nu -1)^2 + 2E=0.
\end{equation}
By solving these equations for $\mu$ and $\nu$ (taking the positive roots) it is obtained that:

\begin{equation}
\mu=\frac{2b}{\sqrt{E+\sqrt{E^2+b^2}}},\qquad  \nu=1-\sqrt{E+\sqrt{E^2+b^2}}.
\label{ParCond}
\end{equation}
The further change of variable $\zeta=2/(1-iz)$ transforms Eq.~(\ref{intermedia}) into
\begin{equation}
\zeta^2\left(1-\zeta \right) \frac {\mbox{d}^2  f}{\mbox{d}\zeta ^2}+
\left[\left(\nu - 2 + i\frac{\mu} 2\right)\zeta^2+2(1-\nu) \zeta \right] \frac {\mbox{d} f}{\mbox{d}\zeta}
+\left[\nu\left(\nu-1\right)-a(a+1)\right] f =0.
\label{cuasiHyp2a}
\end{equation}
By assuming now that $f=\zeta^{\nu+a}\,g$ it turns out that
\begin{equation}
\zeta\left(1-\zeta \right) \frac {\mbox{d}^2  g}{\mbox{d}\zeta ^2}+
\left[ 2(a+1) - \left(\nu+ 2(a+1) -i\frac{\mu}2\right)\zeta   \right] \frac {\mbox{d} g}{\mbox{d}\zeta}
-(\nu+a)\left( a+1-i\frac{\mu}2\right) g =0,
\label{cuasiHyp2}
\end{equation}
which is the hypergeometric equation with parameters $\nu+a$, $a+1-i\frac{\mu} 2$ and $ 2(a+1) $.

Making use of the  Gauss hypergeometric series (see for example \cite{ba53,Abr64,Leb72}), we can construct a set of two linearly independent solutions to Schr\"odinger Eq.~(\ref{StacionaryEq}). It is straightforward to see that the first solutions reads
\begin{eqnarray}\label{solnonconv}
\psi_{L}(x) & = & e^{-[\frac{\mu}2 + i(\nu + a)]x} \sin^{a+1}(x) \,
{}_2F_1\left(\nu  +  a, a  +  1  -  \frac{i\mu}2; 2 a  +  2; 2ie^{-ix}\sin(x)\right). 
\end{eqnarray}
However, except when ${}_2F_1$ becomes a polynomial, the modulus of $2ie^{-ix}\sin(x)$ falls outside the unit circle of convergence of the hypergeometric series in the interval $\pi/6<x<5\pi/6$. In order to solve this issue, let us use the linear transformation formulas of ${}_2F_1$ \cite{Abr64}. Thus, if we employ expressions 15.3.8 and 15.3.3 of \cite{Abr64} it is obtained:
\begin{eqnarray}
\psi_{L}(x)& = & \textstyle 
\kappa_{L}(\mu)e^{-\left[\frac {\mu} 2-i(\nu+a) \right]x}\sin^{a+1} (x) \, {_2}F_1\left(\nu+a, a+1+\frac{i\mu}2;\nu+\frac{i\mu}2; e^{2ix}\right)\nonumber\\[2ex]
&& + \textstyle\rho_{L}(\mu)e^{\left[\frac {\mu} 2+i(1-\nu-a) \right]x}\sin^{-a} (x) \, {_2}F_1\left(1-\nu-a, -a-\frac{i\mu}2;2-\nu-\frac{i\mu}2;e^{2ix}\right),
\label{SolHyp}
\end{eqnarray}
where
\begin{equation*}
\kappa_{L}(\mu)=\textstyle\frac{\Gamma\left( 2a+2\right)\Gamma\left( 1-\nu-i\frac{\mu}2\right)}{\Gamma\left(a+1-i\frac{\mu}2\right)\Gamma\left(a+2-\nu\right)}, \qquad \rho_{L}(\mu)=\textstyle\left(\frac{i}{2}\right)^{2a+1}\frac{\Gamma\left( 2a+2\right)\Gamma\left( \nu-1+i\frac{\mu}2\right)}{\Gamma\left(a+1+i\frac{\mu}2\right)\Gamma\left(a+\nu\right)}.
\end{equation*}
%
This expression does not have already the convergence problem mentioned previously. Moreover, it is straightforward to show that this solutions is real if $E$ is real. 

As we will see below, it is convenient to express the explicit dependence of $\kappa_{L}$ and $\rho_{L}$ on the parameter $\mu$, even though both of them depend as well of the other parameters $a$ and $\nu$. This notation will simplify some of the formulas to be derived next.

In order to obtain from $\psi_{L}(x)$ another linearly independent solution to Schr\"odinger equation~(\ref{StacionaryEq}), let us use the symmetry properties of such equation. Thus, by making a reflexion with respect to $\pi/2$ ($x\rightarrow \pi-x$) and changing the parameter $b$ to $-b$ ($\mu\rightarrow -\mu$), it is obtained once again Eq.~(\ref{StacionaryEq}). However, the associated solution will change, it is denoted as $\psi_{R}(x)$ and it is given by
\begin{eqnarray}
\psi_{R}(x)=  
e^{[\frac{\mu}2 - i(\nu + a)]\pi}e^{-[\frac{\mu}2 - i(\nu + a)]x} \sin^{a+1}(x) \,
{}_2F_1\big(\nu  +  a, a  +  1  +  \frac{i\mu}2; 2 a  +  2; -2ie^{ix}\sin(x)\big)& \nonumber\\
= e^{[\frac{\mu}2 + i(\nu + a)]\pi}
\textstyle 
\kappa_{L}(-\mu)e^{-\left[\frac {\mu} 2+i(\nu+a) \right]x}\sin^{a+1} (x) \, {_2}F_1\left(\nu+a, a+1-\frac{i\mu}2;\nu-\frac{i\mu}2; e^{-2ix}\right)&\nonumber\\
- e^{-[\frac{\mu}2 + i(\nu + a)]\pi}\ \textstyle\rho_{L}(-\mu)e^{\left[\frac {\mu} 2+i(\nu+a-1) \right]x}\sin^{-a} (x) \, {_2}F_1(1-\nu-a, -a+\frac{i\mu}2;2-\nu+\frac{i\mu}2;&\hskip-0.3cme^{-2ix}).
\label{SecSolHyp}
\end{eqnarray}
%
Let us express this formula in terms of the solutions involved in the right hand side of Eq.~(\ref{SolHyp}), by using Eqs.~15.3.7 and 15.3.3 of \cite{Abr64}, in order to obtain:
\begin{eqnarray}
\psi_{R}(x)& = & \textstyle 
\kappa_{R}(\mu)e^{-\left[\frac {\mu} 2-i(\nu+a) \right]x}\sin^{a+1} (x) \, {_2}F_1\left(\nu+a, a+1+\frac{i\mu}2;\nu+\frac{i\mu}2; e^{2ix}\right)\nonumber\\[2ex]
&& + \textstyle\rho_{R}(\mu)e^{\left[\frac {\mu} 2+i(1-\nu-a) \right]x}\sin^{-a} (x) \, {_2}F_1\left(1-\nu-a, -a-\frac{i\mu}2;2-\nu-\frac{i\mu}2;e^{2ix}\right),
\label{SolHyp1}
\end{eqnarray}
where
\begin{eqnarray*}
	\kappa_{R}(\mu)& = & \textstyle\Gamma\left(\nu-i\frac{\mu}2\right)\Gamma\left(1-\nu+i\frac{\mu}2\right)\kappa_{L}(\mu)\left[\frac{e^{\frac{\mu}2\pi}}{\Gamma\left(a+1+i\frac{\mu}2\right)\Gamma\left(-a-i\frac{\mu}2\right)} + \frac{e^{-i\nu\pi}}{\Gamma\left(a+\nu\right)\Gamma\left(1-a-\nu\right)}
	\right], \\
	\rho_{R}(\mu)& = & -\textstyle\Gamma\left(\nu-i\frac{\mu}2\right)\Gamma\left(1-\nu+i\frac{\mu}2\right)\rho_{L}(\mu)\left[\frac{e^{-\frac{\mu}2\pi}}{\Gamma\left(a+1-i\frac{\mu}2\right)\Gamma\left(-a+i\frac{\mu}2\right)} +
	\frac{e^{i\nu\pi}}{\Gamma\left(a+2-\nu\right)\Gamma\left(\nu-a-1\right)}
	\right].
\end{eqnarray*}

Up to here we have obtained a set of two linearly independent solutions  $\{\psi_{L}(x),\psi_{R}(x)\}$ of Eq.~(\ref{StacionaryEq}) vanishing to the left ($x_L=0$) and to the right ($x_R=\pi$), respectively. Thus, the general solution to the Schr\"odinger equation for the TRM potential  (\ref{StacionaryEq}) can be written as follows
\begin{equation}
\psi(x)=A\psi_{L}(x)+B\psi_{R}(x),\qquad A,B \in \setbox0=\hbox{\rm C}\hbox{\hbox
	to0pt{\kern0.4\wd0\vrule height0.9\ht0}\box0}.
\label{GeneralSol}
\end{equation}
The square integrability condition applied to this general solution supplies the physical solutions (bound state eigenfunctions), which are given by (see also Eq.~(\ref{solnonconv})):
%
\begin{equation}
\psi_n(x) = C_n e^{-\left[\frac {b} {n+1+a} -i n \right]x}\sin^{a+1} (x) \, {_2}\textstyle F_1(-n, a+1-\frac {i b} {n+1+a};2a+2;2 i e^{-ix}\sin (x)), 
\label{SolHyp1-1}
\end{equation}
where $C_n$ is a normalization constant. The corresponding energy levels in the spectrum of the TRM Hamiltonian are given by  
\begin{equation}
E_n=\frac 1 2\left(n+a+1\right)^2- \frac {b^2}{2\left(n+a+1\right)^2}, \qquad n \in \rm I\!N.
\label{Spectrum}
\end{equation}
%

Once the general solution~(\ref{GeneralSol}) to the stationary Schr\"odinger equation~(\ref{StacionaryEq}) has been constructed, let us study next some of its properties.

\subsection{Solution analysis}
\label{sol-analysis}
For the purpose of this work let us restrict our analysis to solutions of the form
\begin{equation}
\psi(x;E,\lambda)=\psi_{L}(x)+\lambda\psi_{R}(x),\qquad \lambda=B/A\in \rm I\!R,
\label{GeneralSol2}
\end{equation}
%
where the dependence of the general solution $\psi$ on $E$ and $\lambda$ has been made explicit. From the previous section we realized the importance of knowing the number of zeros for the seed solutions, in order to guarantee the SUSY transformations to be non-singular.

The Sturm's comparison theorem gives us information about the number of zeros for real solutions of Eq.~(\ref{StacionaryEq}), associated to real factorization energies. Thus, given the bound state solutions $\psi_j(x)$, $\psi_{j+1}(x)$ for two consecutive eigenvalues $E_j$, $E_{j+1}$ of $H_0$, a solution with a real factorization energy that is in the gap bounded by $E_j$ and $E_{j+1}$ can have either $j+1$ or $j+2$ nodes. On the other hand, we know that the solutions $\psi_{L}(x)$ and $\psi_{R}(x)$ have the following asymptotic behavior: 
\begin{equation}
\lim_{x \to 0}\psi_{L}(x)=0,\qquad\qquad\left\vert\lim_{x \to \pi}\psi_{L}(x)\right\vert=\infty,
\label{Limit0}
\end{equation}
\begin{equation}
\left\vert\lim_{x \to 0}\psi_{R}(x)\right\vert=\infty,\qquad\qquad\lim_{x \to \pi}\psi_{R}(x)=0,
\label{Limit20}
\end{equation}
%
thus they have $j+1$ nodes for a real factorization energy in the gap $(E_j,E_{j+1})$. Then, it is straightforward to realize that the number of zeros of the solutions (\ref{GeneralSol2}), with factorization energies in $(E_j,E_{j+1})$, depends on the parameter $\lambda$ and the index $j$ of the lower eigenvalue $E_j$. Note as well that $\psi(x;E,\lambda)$ has an even number of nodes if $\lambda>0$ while for $\lambda<0$ it has an odd number of zeros. As a consequence, if $j$ is even and $\lambda>0$ the solution (\ref{GeneralSol2}) has $j+2$ nodes, while for $\lambda\leq 0$ it will have
$j+1$ nodes. On the other hand, for $j$ odd and $\lambda\geq0$ the solution (\ref{GeneralSol2}) has $j+1$ zeros, and $j+2$ nodes for $\lambda<0$. In particular, if the factorization energy is below the ground state energy $E_0$ the solution is nodeless for $\lambda\geq0$ while for $\lambda<0$ it has $1$ node. Moreover, for the two relevant limits of $\lambda$ we get that
\begin{equation}
\lim_{\lambda \to 0}\psi(x;E,\lambda)=\psi_{L}(x), \qquad \lim_{\lambda \to \pm\infty}\psi(x;E,\lambda)\approx\lambda \psi_R(x),
\label{Limitlamda}
\end{equation}
since for $\lambda=0$ the general solution (\ref{GeneralSol2}) reduces to $\psi_L(x)$.

\section{SUSY partners of the TRM potential}
\label{SUSY partners}

Let us take now $V_0(x)$ as the trigonometric Rosen-Morse potential of Eq.~(\ref{tRMp}). The general solution (\ref{GeneralSol2}) to the stationary Schr\"odinger equation for the TRM potential allows us to explore different ways of choosing the factorization energies to implement the SUSY transformations. Depending on this selection, it is posible to generate new Hamiltonians with specific spectral modifications with respect to the initial spectrum. Below we present the results of exploring the three cases of the first-order transformation, as well as several choices of factorization energies and associated seed solutions leading to a final real potential for second-order SUSY \cite{Fer05,Fer10,Fer19}.

\subsection{First-order SUSY partners of $V_0(x)$}
\label{pt1susy}

Let us explore next the three cases discussed at subsect. \ref{1susyqm}.

\subsubsection{Deleting the ground state energy of $H_0$} 
\begin{figure}[h]
	\centering
	\captionsetup[subfloat]{labelfont=bf}
	\begin{tabular}{cc}
		\adjustbox{valign=b}{
			\subfloat[$a=2$,$\quad$ $b=10$, $\quad$ $\epsilon_1=E_0=-1.06$\label{subfig-1:susy1-delete}]{%
				\includegraphics[scale=0.85]{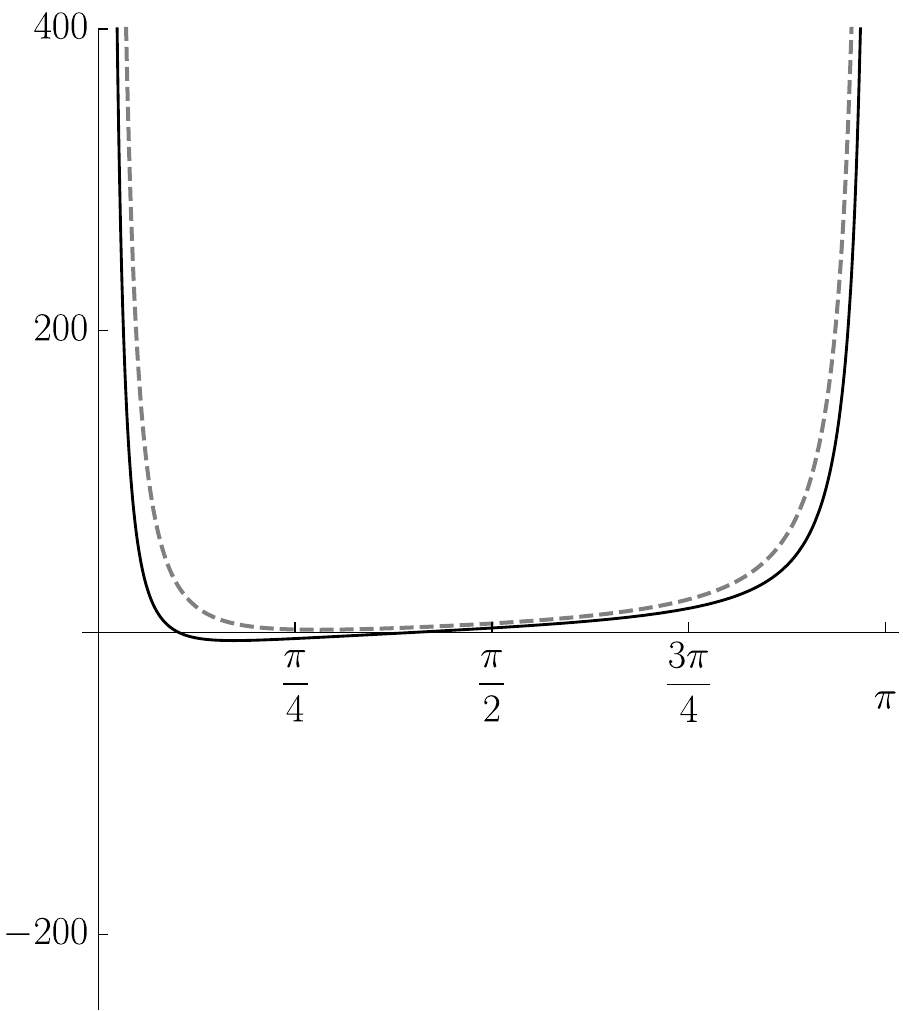}}$\qquad$}
		&      
		\adjustbox{valign=b}{\begin{tabular}{@{}c@{}}
				\subfloat[$a=2$,$\quad$ $b=50$, $\quad$ $\epsilon_1=E_0=-134.39$\label{subfig-2:susy1-delete}]{
					\includegraphics[scale=0.85]{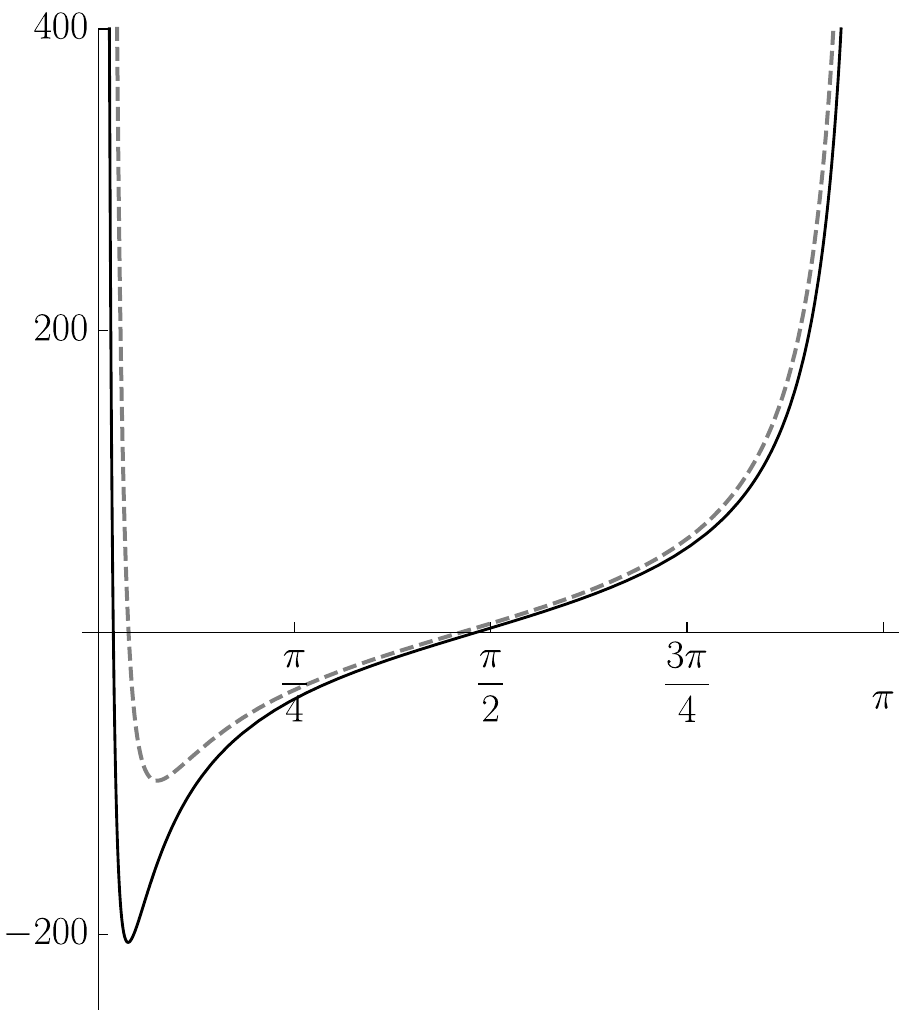}} 
		\end{tabular}}
	\end{tabular}
	\caption{Deleting the ground state. First-order SUSY partners (gray dashed curves) of the TRM potential (black solid curves) for different values of $a$ and $b$. The seed solution $u_{01}(x)$ is the ground state eigenfunction of $H_0$.\label{susy1-delete}}
\end{figure}

In order to implement this transformation let us take as seed solution $u_{01}$ the ground state eigenfunction of $H_0$,
\begin{eqnarray}
&& u_{01}(x)=\psi_{00}(x) \propto \! e^{-\frac{bx}{a+1}} \sin^{a+1}(x) .
\end{eqnarray}
The first-order SUSY partner potential of $V_0(x)$ becomes \cite{ck06} (see Eq.~(\ref{npss})):
\begin{eqnarray}
&& V_1(x) = \frac12(a+1)(a+2)\csc^2(x) - b\cot(x),
\end{eqnarray}
which is once again the Rosen-Morse potential with the parameter change $a\rightarrow a+1$ 
(the well-known shape-invariance of the TRM potential \cite{Bar91,CKS95}). The spectrum of the new Hamiltonian is
\begin{eqnarray}
{\rm Sp}(H_1) = \{E_n, n=1,2,\dots\} .
\end{eqnarray}
An illustration of the SUSY partner potentials $V_0(x)$ and $V_1(x)$ can be seen in Fig.~\ref{susy1-delete}.

\subsubsection{Creating a new ground state} 
\begin{figure}[h]
	\centering
	\captionsetup[subfloat]{labelfont=bf}
	\begin{tabular}{cc}
		\adjustbox{valign=b}{
			\subfloat[\label{subfig-1:susy1-create}]{%
				\includegraphics[scale=.9]{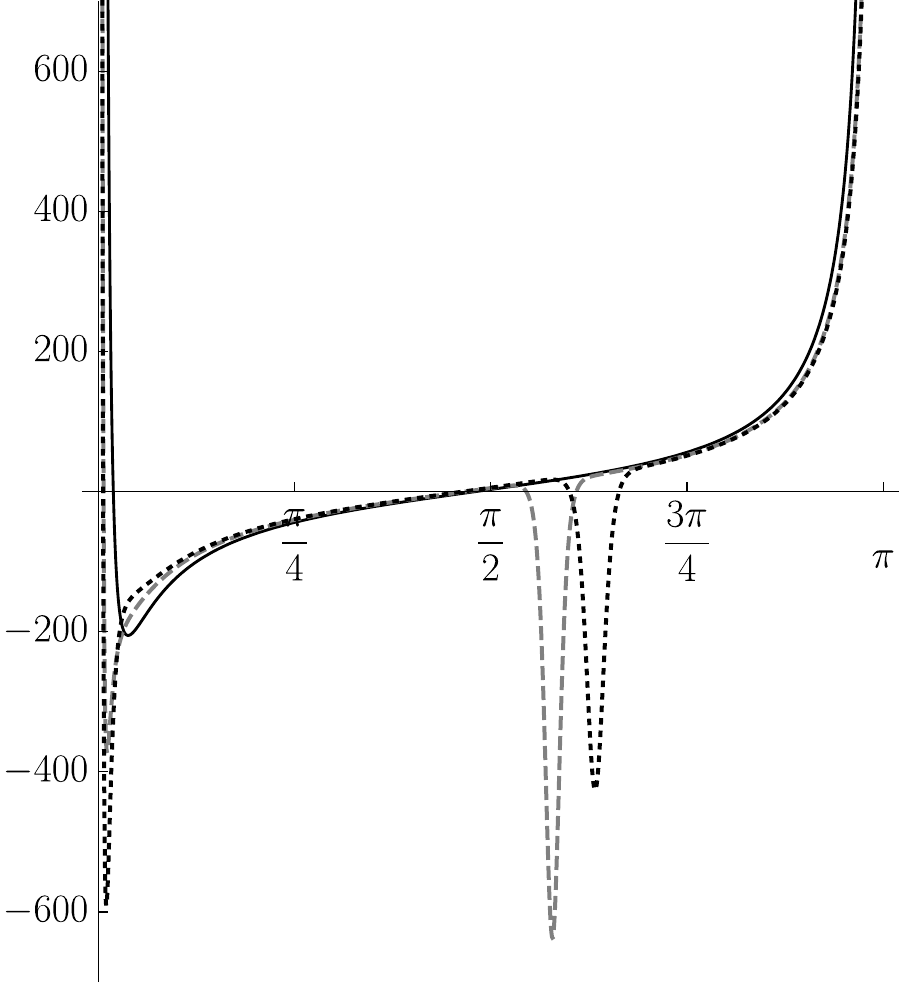}}}
		&      
		\adjustbox{valign=b}{\begin{tabular}{@{}c@{}}
				$\qquad$\subfloat[Seed solution $u_{01}(x)$
				for \mbox{$\epsilon_1=-310.5$}, $\lambda_1=1$ (gray dashed curve) and \mbox{$\epsilon_1=-200$}, $\lambda_1=10$ (black dotted curve).\label{subfig-2:susy1-create}]{%
					\includegraphics[scale=0.62]{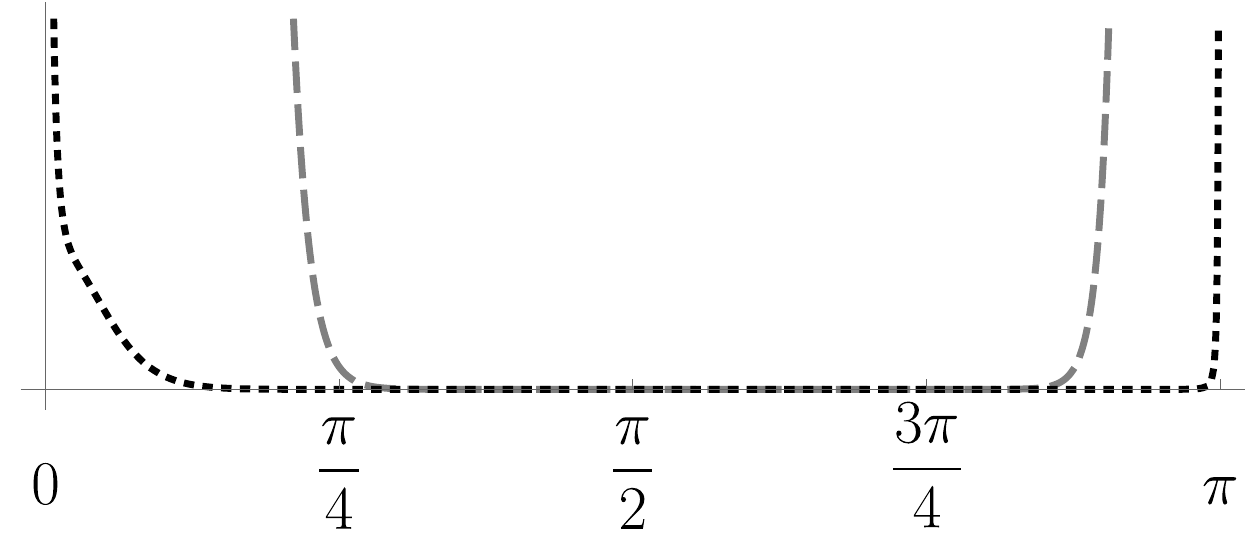}} \\ 
				$\qquad$\subfloat[Missing state eigenfunction $\psi_{1\epsilon_1}(x)$ for \mbox{$\epsilon_1=-310.5,\lambda_1=1$} (gray dashed curve) and \mbox{$\epsilon_1=-200,\lambda_1=10$} (black dotted curve).\label{subfig-3:susy1-create}]{%
					\includegraphics[scale=0.62]{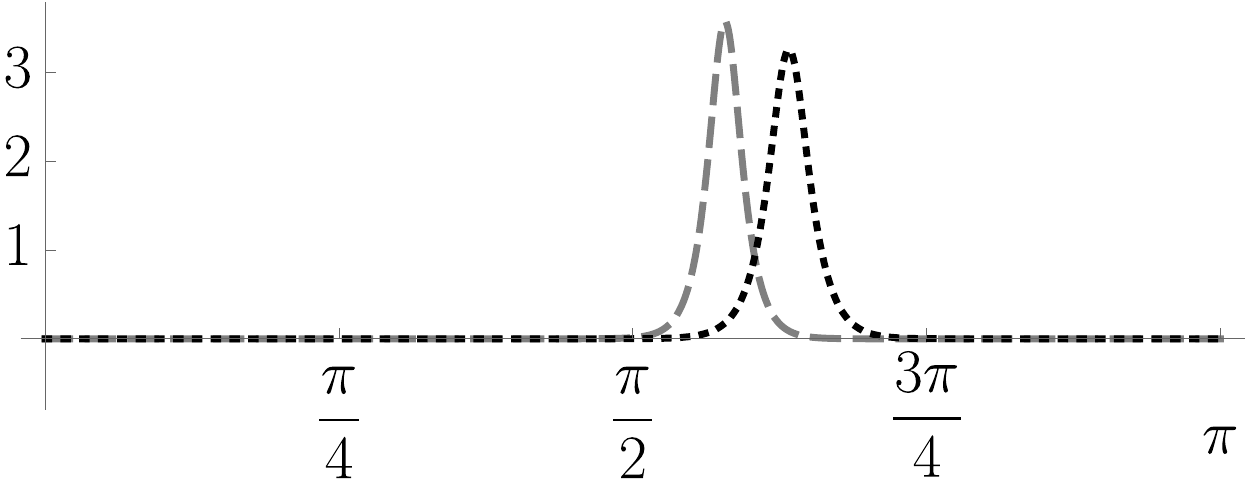}}
		\end{tabular}}
	\end{tabular}
	\caption{Creating a new ground state. \textbf{(a)} TRM potential for $a=2$, $b=50$ (black solid curve) and its first-order SUSY partners for the seed solution $u_{01}(x)$ with $\epsilon_1=-310.5$, $\lambda_1=1$ (gray dashed curve) and $\epsilon_1=-200$, $\lambda_1=10$ (black dotted curve). \textbf{(b)} Seed  solution employed; \textbf{(c)} missing state eigenfunction of $H_1$.\label{susy1-create}}
\end{figure}

In order to generate a new level at $\epsilon_1<E_0$ we need to choose as $u_{01}$ a seed solution (\ref{GeneralSol2}) without nodes and diverging at both ends of the $x$-domain, \textit{i.e.},
\begin{eqnarray}
&& u_{01}(x)= \psi(x,\epsilon_1,\lambda_1)=\sin^{-a}(x) v_{01}(x), \quad \lambda_1>0, \label{ssf1sc}
\end{eqnarray}
which has been factorized appropriately, in order to isolate its divergent behavior at $x=0$ and $x=\pi$ (see also \cite{cf08}). This implies that $v_{01}(x) = \sin^{a}(x) u_{01}(x)$ is a nodeless bounded function in the interval $[0,\pi]$. Thus, the first-order SUSY partner potential of $V_0(x)$ reads
\begin{eqnarray}
&& V_1(x) = \frac12(a-1)a \csc^2(x) - b\cot(x) - [\log v_{01}]''.
\end{eqnarray}
Some plots of potentials belonging to this family are shown in Fig.~\ref{susy1-create}.

\subsubsection{Isospectral transformations} 
\begin{figure}[h]
	\centering
	\captionsetup[subfloat]{labelfont=bf}
	\begin{tabular}{cc}
		\adjustbox{valign=b}{
			\subfloat[\label{subfig-1:susy1-isospectral}]{%
				\includegraphics[scale=.9]{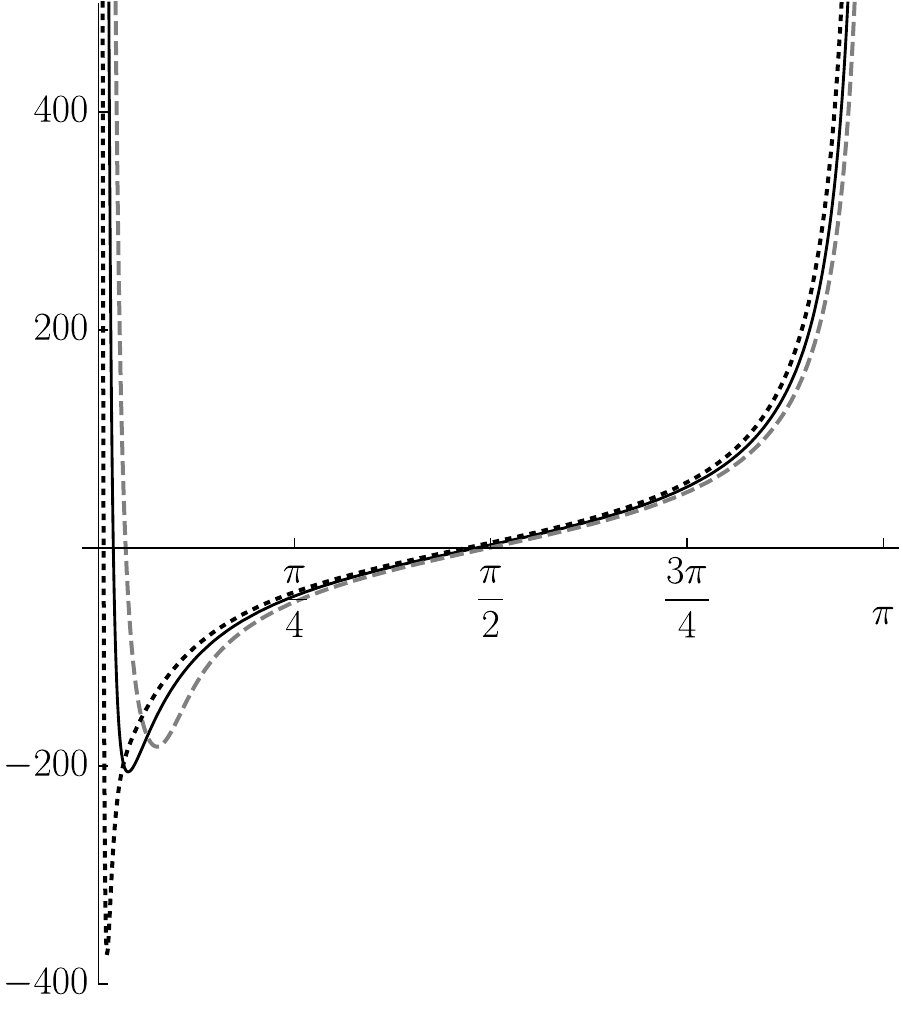}}}
		&      
		\adjustbox{valign=b}{\begin{tabular}{@{}c@{}}
				$\qquad$\subfloat[Seed solution \mbox{$u_{01}(x) = \psi_{L}(x)$} for \mbox{$\epsilon_1=-310.5$} (gray dashed curve) and \mbox{$u_{01}(x) = \psi_{R}(x)$ for \mbox{$\epsilon_1=-200$}} (black dotted curve).\label{subfig-2:susy1-isospectral}]{%
					\includegraphics[scale=0.625]{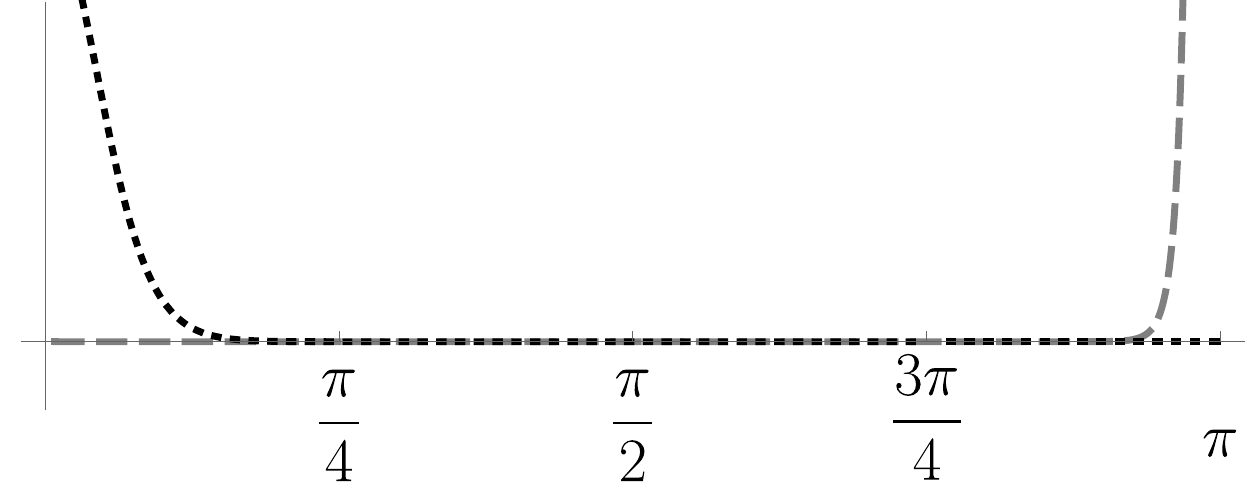}} \\
				$\qquad$\subfloat[Missing state wavefunction \mbox{$\psi_{1\epsilon_1}(x)$} for \mbox{$\epsilon_1=-310.5$} (gray dashed curve) and \mbox{$\epsilon_1=-200$} (black dotted curve).\label{subfig-3:susy1-isospectral}]{%
					\includegraphics[scale=0.625]{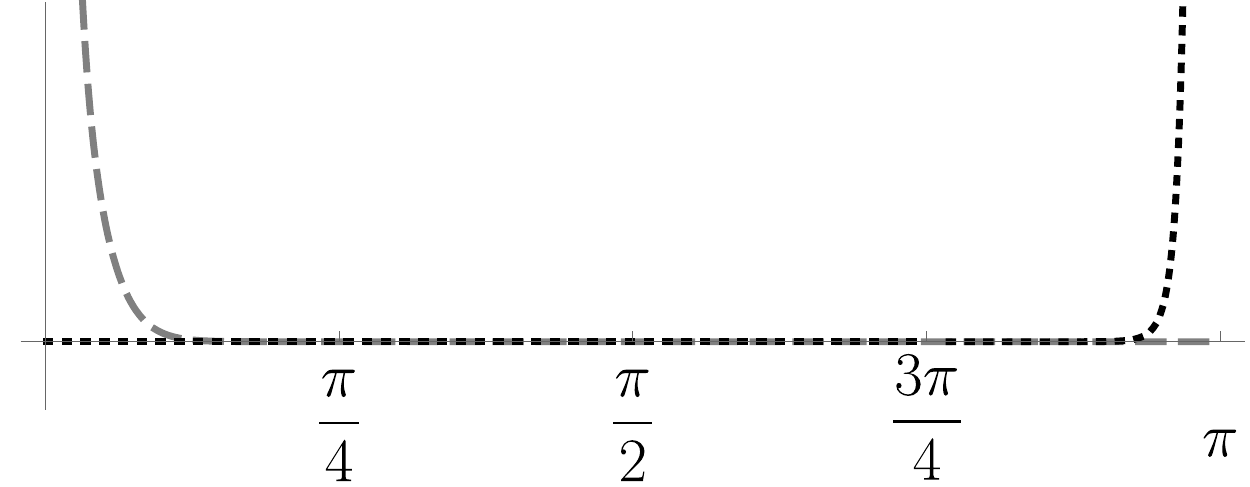}}
		\end{tabular}}
	\end{tabular}
	\caption{	Isospectral transformation. \textbf{(a)} TRM potential for $a=2$, $b=50$ (black solid curve) and its first-order SUSY partners for $\epsilon_1=-200$ (gray dashed curve) and $\epsilon_1=-310.5$ (black dotted curve). \textbf{(b)} Seed solution employed;  \textbf{(c)} missing state wavefunction.\label{susy1-isospectral}}
\end{figure}

They are obtained from the previous case, either as the limit when $\lambda_1\rightarrow 0$ ($u_{01} = \psi_{L}(x)$) or when $\lambda_1\rightarrow \infty$ ($u_{01} \approx\lambda_1\psi_{R}(x)$). In both limits we obtain that
\begin{eqnarray}
&& V_1(x) = \frac12a(a+1)\csc^2(x) - b\cot(x) - [\log u_{01}]'' .
\end{eqnarray}
This time we did not factorize the seed solution as in Eq.~(\ref{ssf1sc}), since now its asymptotic behaviors at $x=0$ and $x=\pi$ are different, thus inducing opposite changes in the coefficient of the term $\csc^2(x)$ at both ends of the $x$-domain ($a\rightarrow a+1$ at one end and $a\rightarrow a-1$ at the other). An illustration of this potential is shown in Fig.~\ref{susy1-isospectral}.

\subsection{Second-order SUSY partners of $V_0(x)$}
\label{pt2susy}

Let us explore now the second-order SUSY transformations of subsect. \ref{2susyqm}. In the three cases of the classification presented there, in general, the second-order SUSY partner potentials $V_2(x)$ of the trigonometric Rosen-Morse potential take the form
\begin{eqnarray}
&& V_2(x) = \frac12 a(a+1) \csc^2(x) - b \cot (x) - \{\ln[W(u_{01},u_{02})]\}'' . \label{V2gen}
\end{eqnarray}
Sometimes we could get more convenient expressions; they will be reported when this happens.

\subsubsection{Real case $(c>0)$}

There are several kinds of modified spectrum for $H_2$ when two real different factorization energies are taken.

\begin{itemize}
	
	\item[(i)] Deleting two energy levels 
	
	This spectral modification is achieved if the factorization energies are chosen as two consecutive eigenvalues of $H_0$, $\epsilon_2=E_j, \ \epsilon_1=E_{j+1}$. The seed solutions employed are the two physical eigenfunctions with such energies, $u_{01}(x) = \psi_{0 j+1}(x)$, $u_{02}(x) = \psi_{0j}(x)$, which vanish at the two ends $x=0$ and $x=\pi$ of the potential domain. Taking into account now Eq.~(\ref{newEpFreal}), which relates $u_{01}, \, u_{02}$ with the two formal eigenfunctions of $H_2$ associated to $E_j, \ E_{j+1}$, it turns out that $\psi_{2\epsilon_1}, \, \psi_{2\epsilon_2}$ do not fulfill the boundary conditions, thus $E_j, \ E_{j+1}$ do not belong to the spectrum of $H_2$. One concludes that the effective action of this transformation is to delete $E_j, \ E_{j+1}$ for generating $H_2$.
	
	Let us note that this particular transformations has been studied previously \cite{Dom11}, the resulting potentials are given by 
	\begin{eqnarray}
	&& V_2(x) =\frac12 (a+2)(a+3) \csc^2(x) -b \cot (x)- \left[\ln\left({\cal W}(x)\right)\right]'',
	\label{tRMrealdelete2}
	\end{eqnarray}
	where ${\cal W}(x)$ is the part of the Wronskian which is left once the null behavior of $W(u_{01},u_{02})$ at both ends has been isolated, \textit{i.e.}, 
	\begin{align*}
	W(u_{01}(x),u_{02}(x))=\sin^{2a+3}(x){\cal W}(x).\nonumber
	\end{align*}
	%
	Note that with this factorization the singularities induced by $W(u_{01},u_{02})$ on the new potential at the end points can be separated, in order to group them with the singularities of the TRM potential (see also \cite{cf08}).\\
	
	A special case emerges for $j=0$, when the two seed solutions are the ground state and the first excited state of $H_0$. Now a shape invariant SUSY partner potential appears, \textit{i.e.}, the new potential becomes the initial one with the parameter $a$ increased by two. In this case the last term in Eq.~(\ref{tRMrealdelete2}) vanishes, which corresponds to substitute $a$ by $a+2$ in the initial potential in order to obtain $V_2(x)$.
	
	In Fig.~\ref{subfig-1:real-case-erasing} we can see some potentials generated by this transformation, while the corresponding seed solutions are shown in Figs.~\ref{subfig-2:real-case-erasing} and \ref{subfig-3:real-case-erasing}.\\
	
	\begin{figure}[h]
		\centering
		\captionsetup[subfloat]{labelfont=bf}
		\begin{tabular}{cc}
			\adjustbox{valign=b}{
				\subfloat[\label{subfig-1:real-case-erasing}]{%
					\includegraphics[scale=.9]{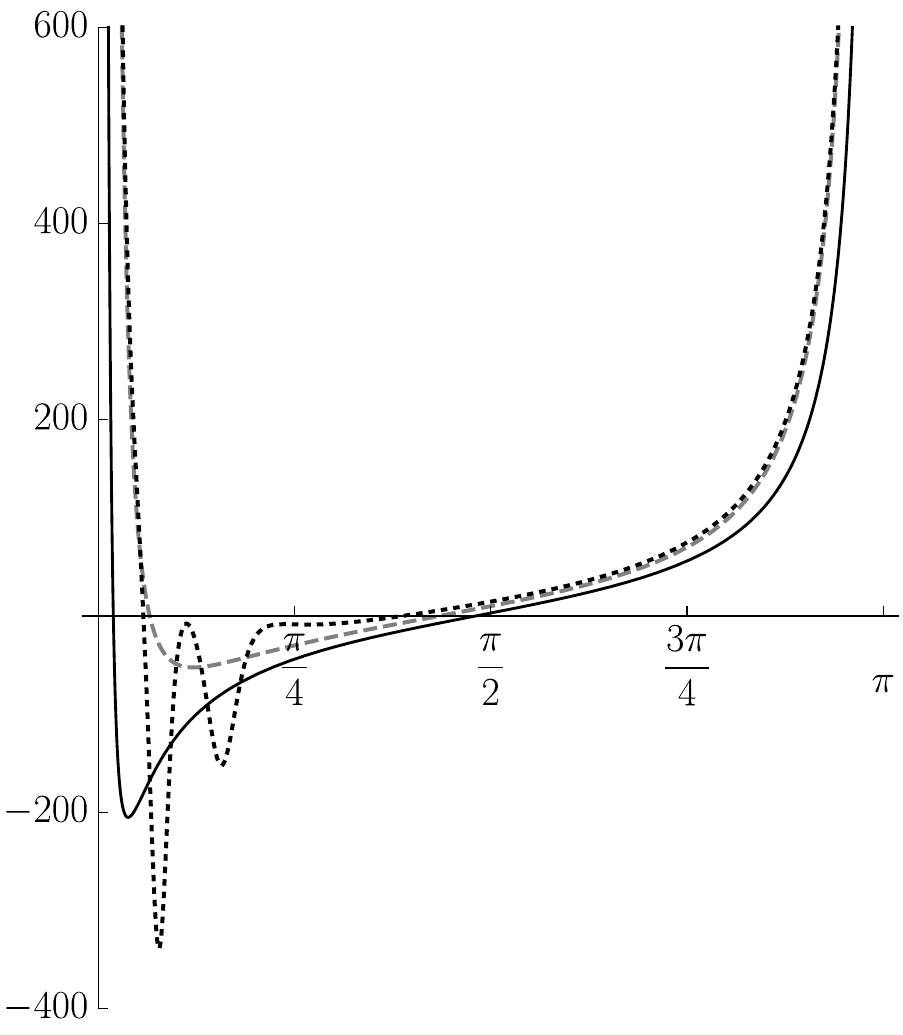}}}
			&      
			\adjustbox{valign=b}{\begin{tabular}{@{}c@{}}
					$\qquad$\subfloat[Seed solutions $u_{01}(x) = \psi_{01}(x)$ (black curve) and $u_{02}(x)= \psi_{00}(x)$ (gray curve).\label{subfig-2:real-case-erasing}]{%
						\includegraphics[scale=0.615]{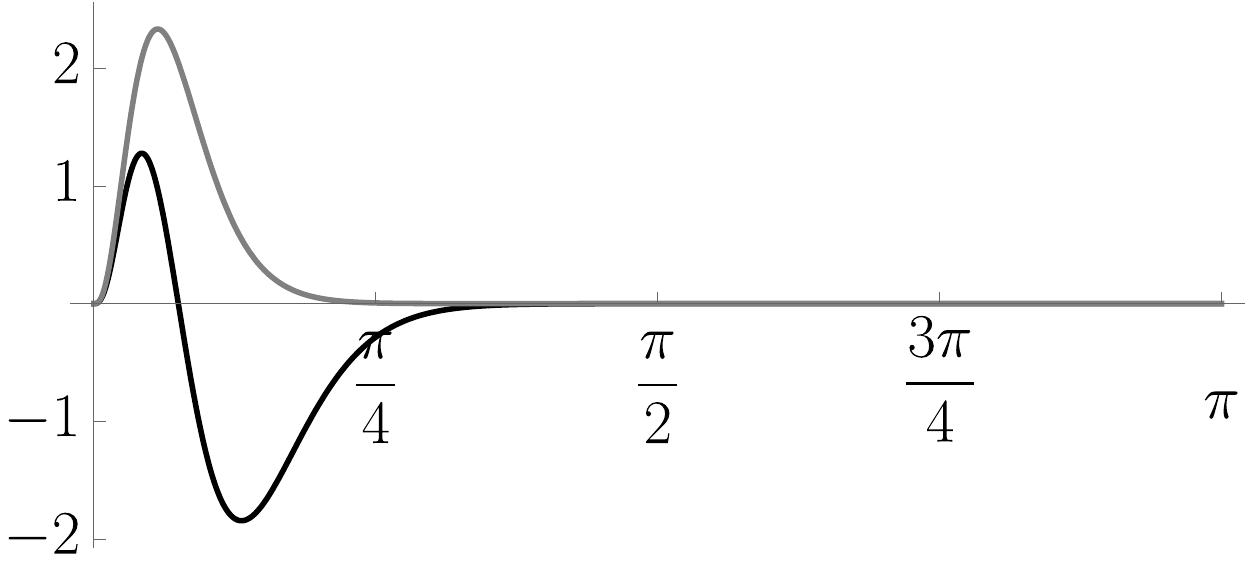}} \\ 
					$\qquad$\subfloat[Seed solutions $u_{01}(x)= \psi_{03}(x)$ (black curve) and $u_{02} (x)= \psi_{02}(x)$ (gray curve).\label{subfig-3:real-case-erasing}]{%
						\includegraphics[scale=0.615]{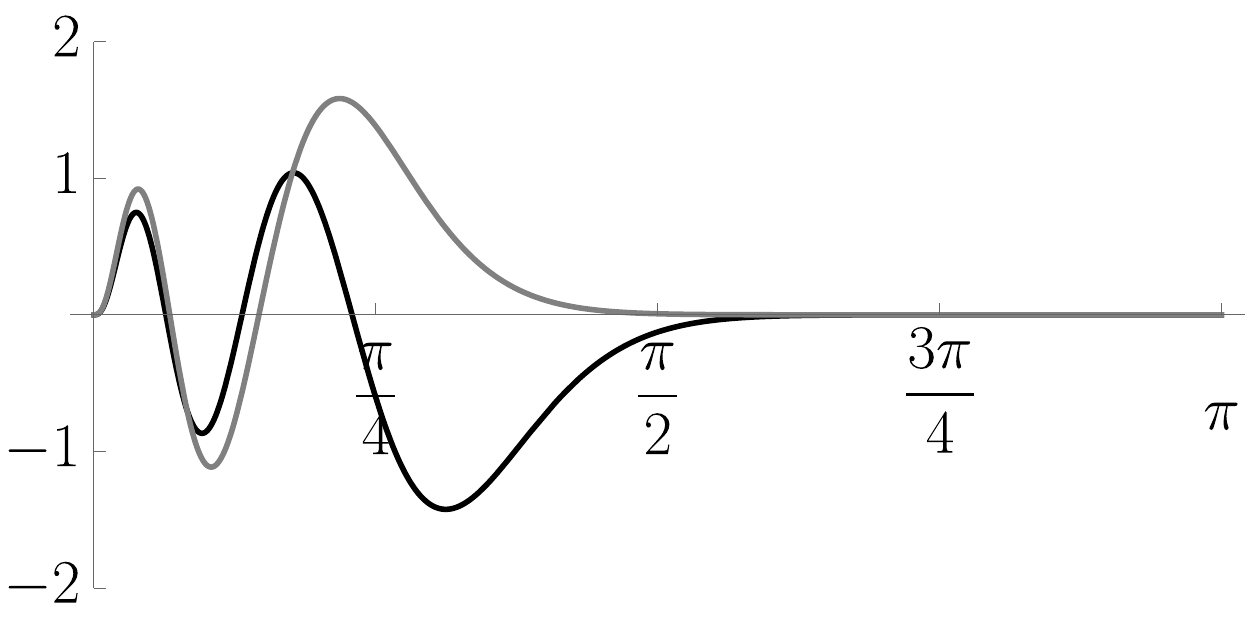}}
			\end{tabular}}
		\end{tabular}
		\caption{Deleting two energy levels. \textbf{(a)} TRM potential for $a=2$, $b=50$ (black solid curve) and its second-order SUSY partners $V_2(x)$ for $\epsilon_1=E_1=-70.125$, $\epsilon_2=E_0=-134.389$ (gray dashed curve) and $\epsilon_1=E_3=-16.7222$, $\epsilon_2=E_2=-37.5$ (black dotted curve). Seed solutions used to plot the gray dashed curve \textbf{(b)} and the black dotted curve \textbf{(c)} for $V_2(x)$.\label{real-case-erasing}}
	\end{figure}

	\item[(ii)] Creating two energy levels
	
	Let us choose the factorization energies such that $E_j<\epsilon_2<\epsilon_1<E_{j+1}$. We classify first the energy gap $(E_{j},E_{j+1})$ into two types, according to either $j$ is even or odd, in order to indicate then the way of choosing the solutions with the right number of zeros for the  transformation to be non-singular. If $j$ is even we need to take $\lambda_1<0$ in the seed solution $u_{01}(x)=\psi(x;\epsilon_1,\lambda_1)$ (with $j+1$ nodes) while a $\lambda_2>0$ has to be chosen in $u_{02}(x)=\psi(x;\epsilon_2,\lambda_2)$ (with $j+2$ nodes). However, if $j$ is odd this $\lambda$-choice for each seed solution should be reversed. Note that the seed solutions employed are non-physical, but the transformation produces physical solutions (eigenfunctions) of $H_2$ for the new eigenvalues $\epsilon_1$ and $\epsilon_2$.
	
	The potentials resulting of this transformation take the form
	\begin{eqnarray}
	&& V_2(x) =\frac12 (a-2)(a-1) \csc^2(x) -b \cot (x)- \left[\ln\left({\cal{W}}(x)\right)\right]'',
	\label{tRMrealcreate2}
	\end{eqnarray}
	where ${\cal {W}}(x)$ is now the nodeless bounded function which appears once the divergent behavior induced by the seed solutions on the Wronskian at $x=0$ and $x=\pi$ is isolated, namely, 
	\begin{align*}
	W(u_{01}(x),u_{02}(x))=\sin^{-2a+1}(x){\cal{W}}(x).
	\end{align*}
	
	The lowest energy gap $(-\infty,E_0)$ is worth of some discussion. For seed solutions such that $\epsilon_2<\epsilon_1<E_0$ we need to choose $\lambda_1>0$ and $\lambda_2<0$. This case is equivalent to make two successive non-singular first-order transformations, the first one with factorization energy $\epsilon_1$ and the second one with $\epsilon_2$.
	
	In Fig.~\ref{subfig-1:real-case-creating}  we can see plots of the potentials generated by this transformation for $\epsilon_2<\epsilon_1<E_0$ (gray dashed curve) and $E_3<\epsilon_2<\epsilon_1<E_4$ (black dotted curve). The seed solutions employed are shown to the right, in Fig.~\ref{subfig-2:real-case-creating} for the potential plotted in the gray dashed curve and in Fig.~\ref{subfig-3:real-case-creating} for the one in the black dotted curve.\\

	\begin{figure}[h]
		\centering
		\captionsetup[subfloat]{labelfont=bf}
		\begin{tabular}{cc}
			\adjustbox{valign=b}{
				\subfloat[\label{subfig-1:real-case-creating}]{%
					\includegraphics[scale=.9]{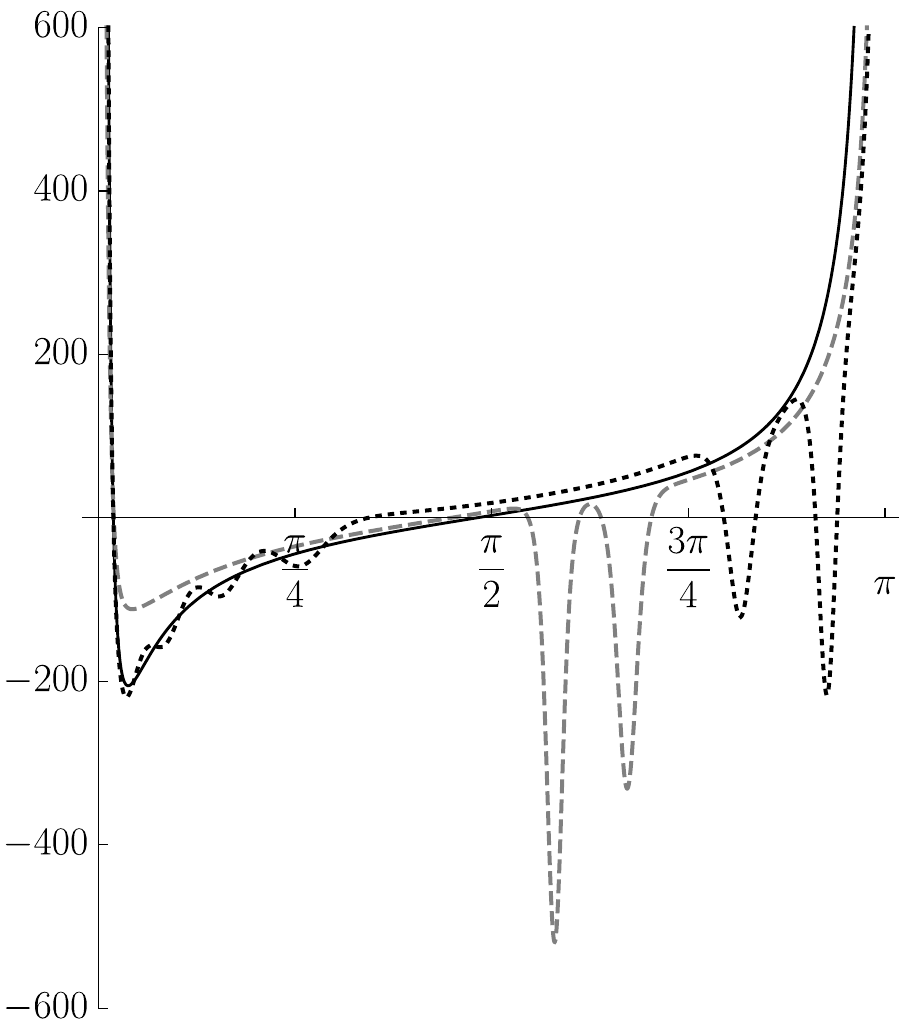}}}
			&      
			\adjustbox{valign=b}{\begin{tabular}{@{}c@{}}
					$\qquad$\subfloat[Seed solutions \mbox{$u_{01}(x)$} for \mbox{$\epsilon_1=-150$}, \mbox{$\lambda_1=1$} (gray curve) and \mbox{$u_{02}(x)$} for \mbox{$\epsilon_2=-250$}, \mbox{$\lambda_2=-1$} (black curve).\label{subfig-2:real-case-creating}]{%
						\includegraphics[scale=0.62]{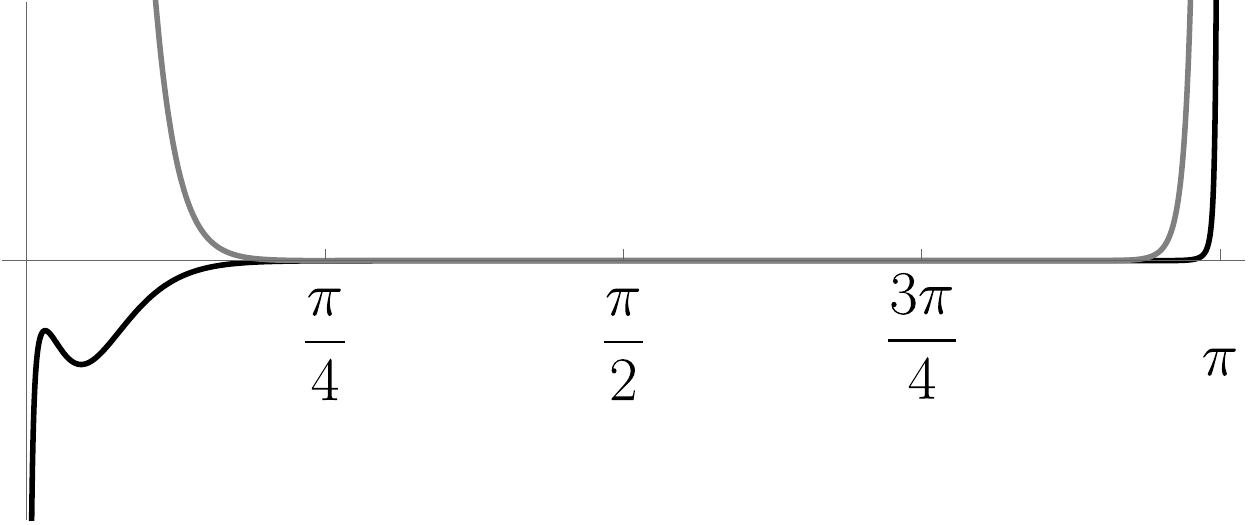}} \\
					$\qquad$\subfloat[Seed solutions \mbox{$u_{01}(x)$} for \mbox{$\epsilon_1=-2$}, \mbox{$\lambda_1=1$} (gray curve) and \mbox{$u_{02}(x)$} for \mbox{$\epsilon_2=-10$}, \mbox{$\lambda_2=-1$} (black curve).\label{subfig-3:real-case-creating}]{%
						\includegraphics[scale=0.62]{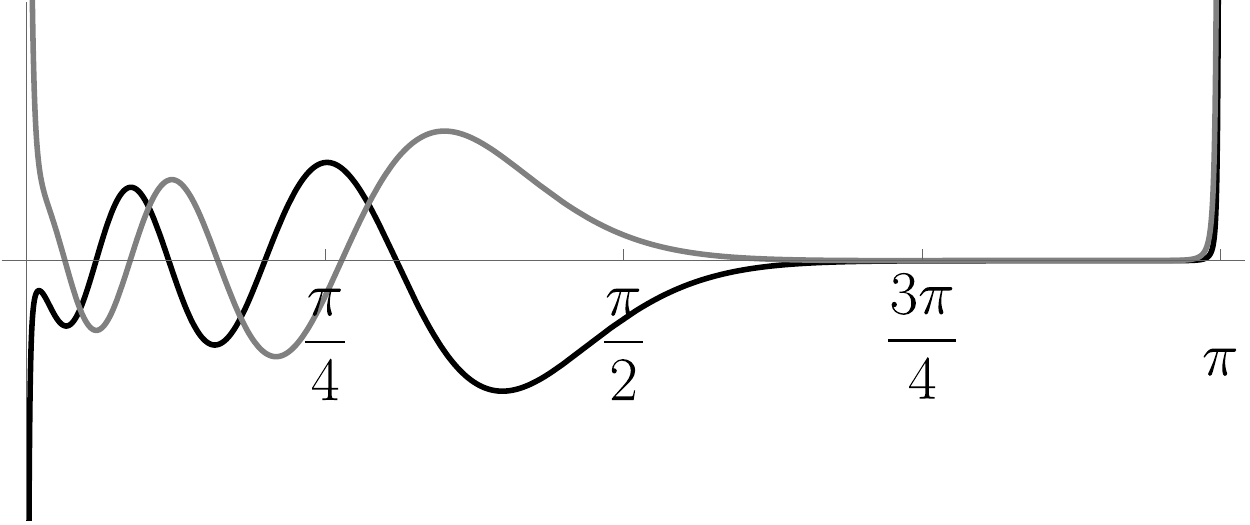}}
			\end{tabular}}
		\end{tabular}
		\caption{Creating two energy levels. \textbf{(a)} TRM potential for $a=2$, $b=50$ (black solid curve) and its second-order SUSY partners $V_2(x)$ for $\epsilon_1=-150$, $\epsilon_2=-250$ (gray dashed curve) and $\epsilon_1=-2$, $\epsilon_2=-10$ (black dotted curve). Seed solutions used to plot the gray dashed curve \textbf{(b)} and the black dotted curve \textbf{(c)} of $V_2(x)$.\label{real-case-creating}}
	\end{figure}
	
	\item[(iii)] Creating one energy level
	
	This spectral change is achieved as a limit of the previous case (ii), when one of the non-physical seed solutions becomes either $\psi_L(x)$ or $\psi_R(x)$ (for $\lambda\rightarrow 0$ or $\lambda\rightarrow \pm\infty$, respectively). If $E_j<\epsilon_2<\epsilon_1<E_{j+1}$ and the seed solution for $\epsilon_1$ is either $\psi_L(x)$ or $\psi_R(x)$, then we have to take care that the transformation function for $\epsilon_2$ should have $j+2$ nodes ($\lambda_2>0$ for $j$ even or $\lambda_2<0$ for $j$ odd). On the other hand, if the seed solution for $\epsilon_2$ is either $\psi_L(x)$ or $\psi_R(x)$, then the transformation function for $\epsilon_1$ should have $j+1$ nodes ($\lambda_1<0$ if $j$ is even or $\lambda_1>0$ if $j$ is odd).
	
	The new potential $V_2(x)$ is given by Eq.~(\ref{V2gen}). Now a change in the coefficient of the singular term $\csc^2(x)$ arises ($a\rightarrow a-2$) at the end where both $u_{0i}(x)$ diverge while there is no change at the other end, where one solution diverges but the other vanishes.
	
	The energy gap $(-\infty,E_0)$ is again worth of consideration. In this case if the seed solution for $\epsilon_1$ is of type $\psi_R(x)$ or $\psi_L(x)$ then the transformation function for $\epsilon_2$ should have $\lambda_2<0$, while if the seed solution for $\epsilon_2$ is either of type $\psi_R(x)$ or $\psi_L(x)$ then the one for $\epsilon_1$ should have $\lambda_1>0$.
	
	In Fig.~\ref{subfig-1:real-case-createone} we can see plots of the potentials generated by this transformation, which create a new level at $\epsilon_2=-60$ (gray dashed curve) and at $\epsilon_2=-10$ (black dotted curve), while the seed solutions used to implement such transformations can be seen in Figs.~\ref{subfig-2:real-case-createone} and \ref{subfig-3:real-case-createone}.\\
	
	\begin{figure}[h]
		\centering
		\captionsetup[subfloat]{labelfont=bf}
		\begin{tabular}{cc}
			\adjustbox{valign=b}{
				\subfloat[\label{subfig-1:real-case-createone}]{
					\includegraphics[scale=.9]{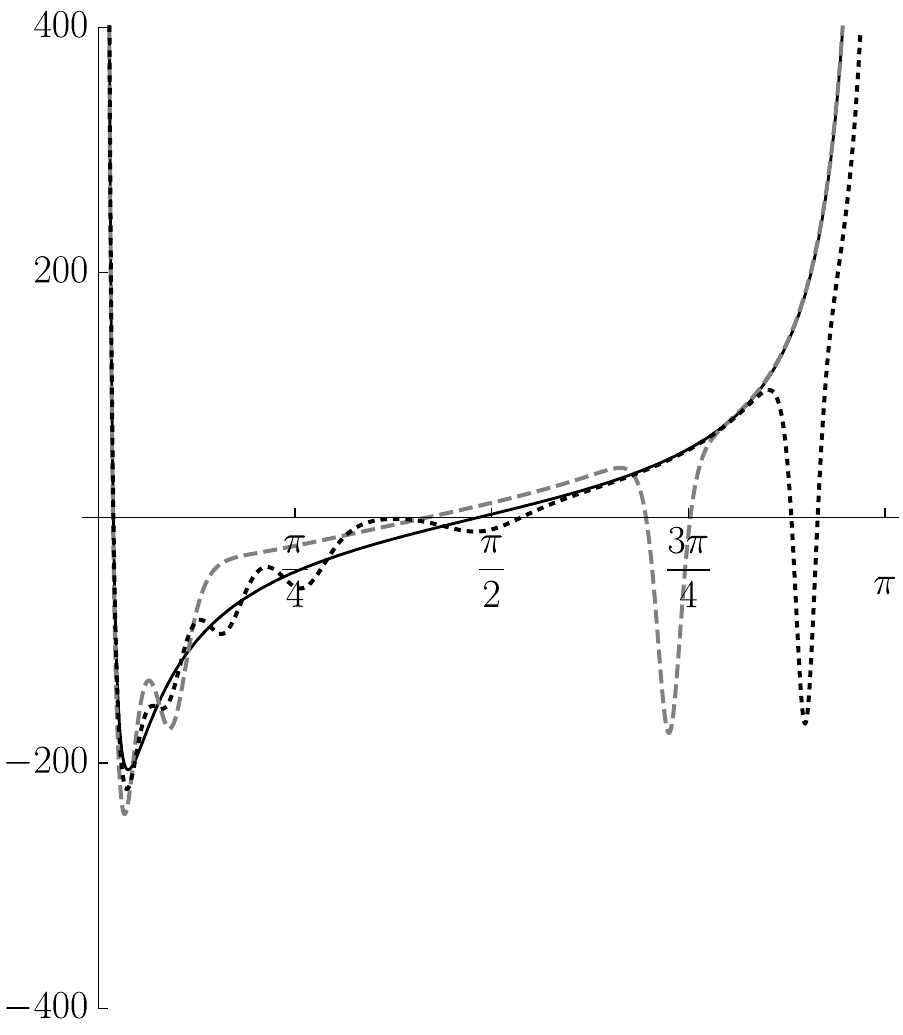}}}
			&      
			\adjustbox{valign=b}{\begin{tabular}{@{}c@{}}
					$\qquad$\subfloat[Seed solutions \mbox{$u_{01}(x)=\psi_R(x)$} for \mbox{$\epsilon_1=-40$} (black curve) and \mbox{$u_{02}(x)$} for \mbox{$\epsilon_2=-60$}, $\lambda_2=-1$ (gray curve).\label{subfig-2:real-case-createone}]{%
						\includegraphics[scale=0.62]{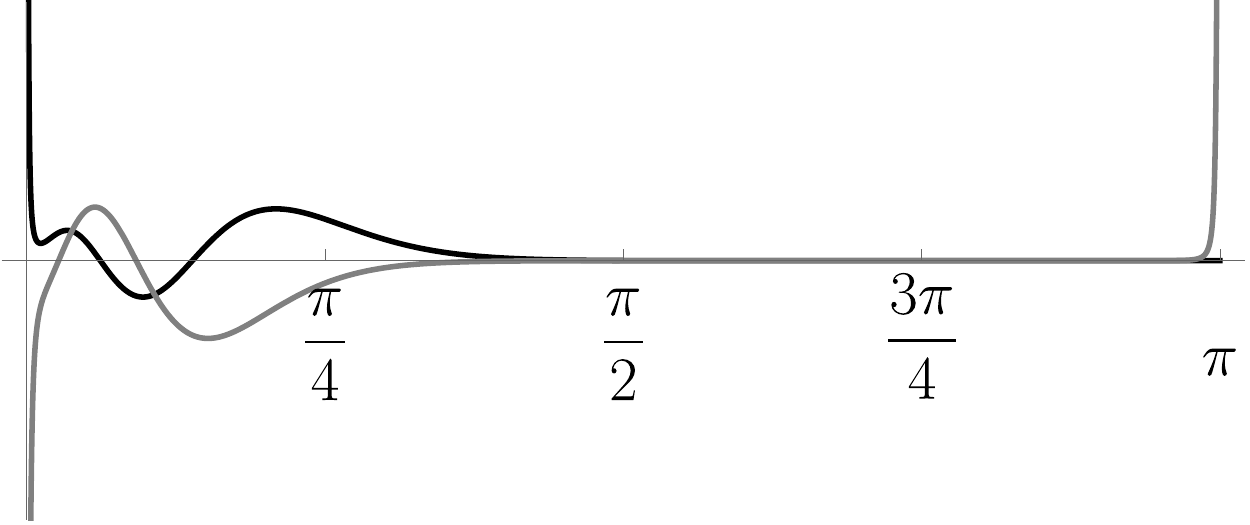}} \\ 
					$\qquad$\subfloat[Seed solutions \mbox{$u_{01}(x)=\psi_L (x)$} for \mbox{$\epsilon_1=-2$} (black curve) and  
					\mbox{$u_{02}(x)$} for \mbox{$\epsilon_2=-10$}, $\lambda_2=-1$ (gray curve).\label{subfig-3:real-case-createone}]{%
						\includegraphics[scale=0.62]{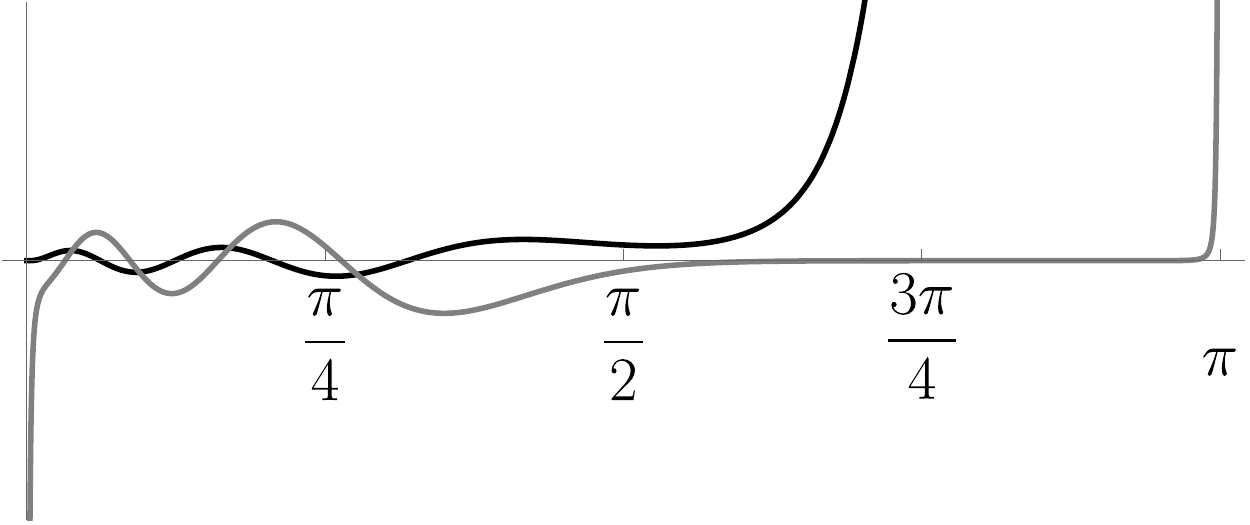}}
			\end{tabular}}
		\end{tabular}
		\caption{Creating one energy level. \textbf{(a)} TRM potential for $a=2$, $b=50$ (black solid curve) and its second-order SUSY partners $V_2(x)$ for $\epsilon_1=-40$, $\epsilon_2=-60$ (gray dashed curve) and $\epsilon_1=-2$, $\epsilon_2=-10$ (black dotted curve). Seed solutions used to plot the gray dashed curve \textbf{(b)} and the black dotted curve \textbf{(c)} of $V_2(x)$.\label{real-case-createone}}
	\end{figure}	
	
	\item[(iv)] Moving one energy level
	
	This kind of spectral modification requires to mix the two types of factorization energies already discussed, deleting one arbitrary level $E_j$ of the initial spectrum but creating simultaneously the new level at a factorization energy which is in its neighbor energy gap, with an associated non-physical seed solution. 
	
	The domain of $\lambda$ for the non-physical seed solution to induce non-singular transformations depends on either its factorization energy is above or below the mobile level. Thus, when the factorization energies are taken as $\epsilon_1=E_j$ and $E_{j-1}<\epsilon_2<E_j$ the level $E_j$ is moved down, but for $\epsilon_2=E_j$ and $E_{j}<\epsilon_1<E_{j+1}$ the level $E_j$ is moved up. In both cases the non-physical seed solution must have $j+1$ nodes, which implies that if $j$ is even $\lambda<0$ while if $j$ is odd $\lambda>0$.
	
	As there are not singularities induced by the Wronskian at $x=0$ and $x=\pi$, the second-order SUSY partner potential $V_2(x)$ is given by Eq.~(\ref{V2gen}). This is so since the singularity induced by the null behavior of the bound state is canceled out by the singularity produced by the divergent behavior of the non-physical seed solution at both ends of the $x$-domain, thus the term $\csc^2(x)$ in $V_2(x)$ keeps unaltered, \textit{i.e.}, without changing the coefficient $a(a+1)/2$.
	
	In Fig.~\ref{subfig-1:real-case-moving} we can see plots of the potentials for this case, the dashed curve for the potential that results of moving down the level $E_1$, the dotted curve for the potential generated by moving up the level $E_4$. The seed solutions used to implement the transformation are shown in Figs.~\ref{subfig-2:real-case-moving} and \ref{subfig-3:real-case-moving}.\\
	
	\begin{figure}[h]
		\centering
		\captionsetup[subfloat]{labelfont=bf}
		\begin{tabular}{cc}
			\adjustbox{valign=b}{
				\subfloat[\label{subfig-1:real-case-moving}]{%
					\includegraphics[scale=.9]{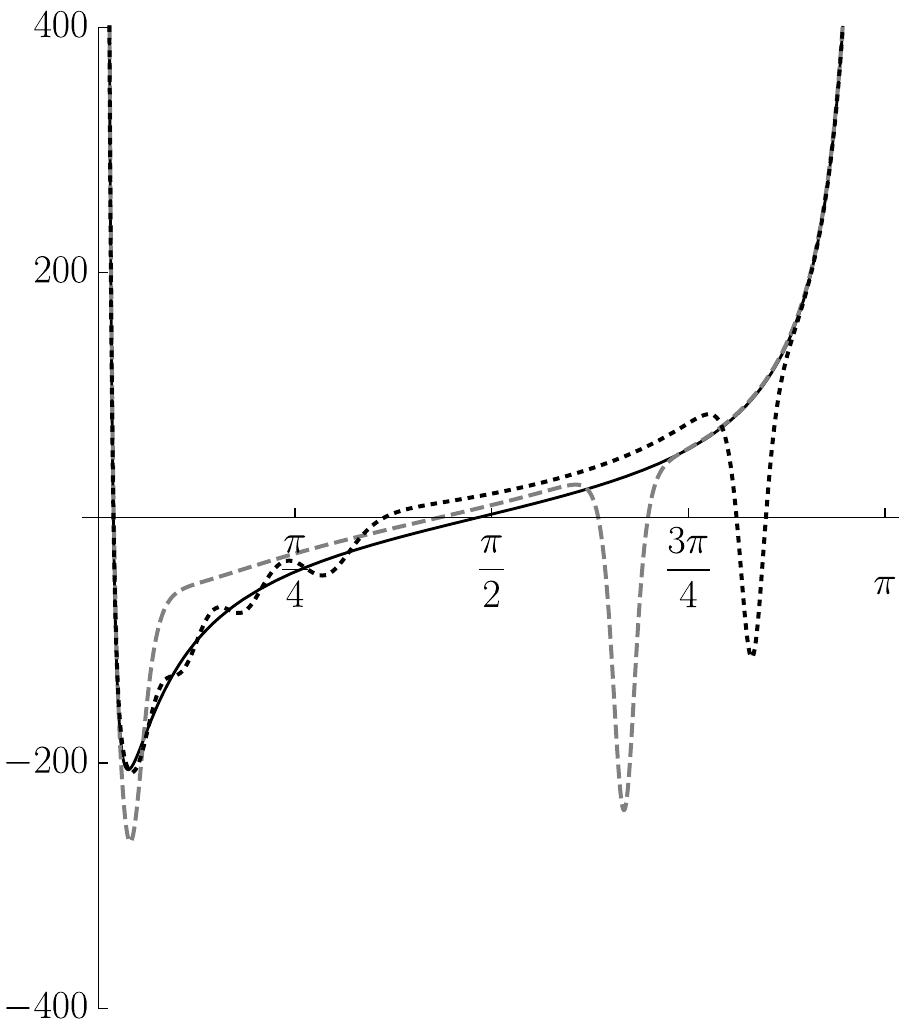}}}
			&      
			\adjustbox{valign=b}{\begin{tabular}{@{}c@{}}
					$\qquad$\subfloat[Seed solutions $u_{01}(x)=\psi_{01}(x)$ (black curve) and \mbox{$u_{02}(x)$} for \mbox{$\epsilon_2=-100$}, $\lambda_2=1$ (gray curve).\label{subfig-2:real-case-moving}]{%
						\includegraphics[scale=.62]{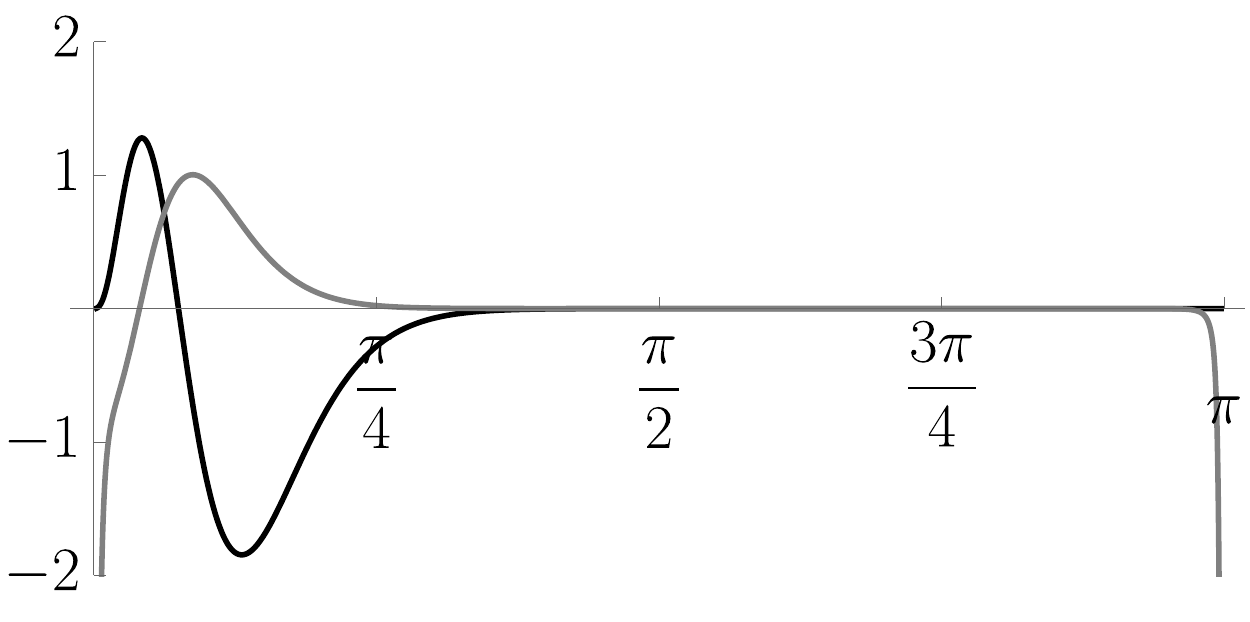}} \\
					$\qquad$\subfloat[Seed solutions 
					\mbox{$u_{01}(x)$} for \mbox{$\epsilon_1=0$}, \mbox{$\lambda_1=-1$} (black curve) and $u_{02}(x)=\psi_{04}(x)$ (gray curve).\label{subfig-3:real-case-moving}]{%
						\includegraphics[scale=.62]{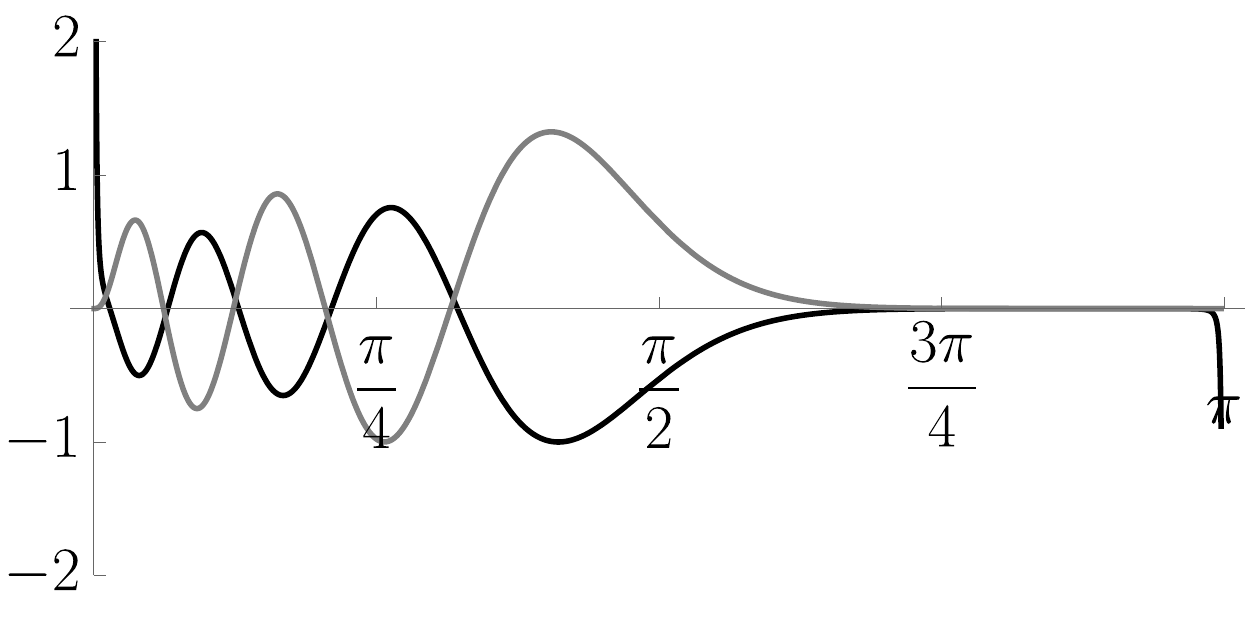}}
			\end{tabular}}
		\end{tabular}
		\caption{Moving one energy level. \textbf{(a)} TRM potential for $a=2$, $b=50$ (black solid curve) and its second-order SUSY partners $V_2(x)$ for $\epsilon_1=E_1=-70.125$, $\epsilon_2=-100$ (gray dashed curve) and for $\epsilon_1=0$, $\epsilon_2=E_4=-1.01$ (black dotted curve). Seed solutions used to plot the gray dashed curve \textbf{(b)} and the black dotted curve \textbf{(c)} of $V_2(x)$.\label{real-case-moving}}
	\end{figure}

	\item[(v)] Deleting one energy level 
	
	Another possible spectral change is to delete $E_j$ from the initial spectrum. This is obtained as a limit of case (iv), when the eigenfunction of $H_2$ associated to the new energy level leaves to be square-integrable but the transformation is still non-singular. Thus, one factorization energy must be $E_j$, with the seed solution being taken as $\psi_{0j}$. The other factorization energy must be in the domain $(E_{j-1},E_j)\cup(E_j,E_{j+1})$, with the seed solution chosen as $\psi_{R}(x)$ or $\psi_{L}(x)$, which is obtained by making either $\lambda\rightarrow 0$ or $\lambda\rightarrow \pm\infty$ in the general solution (\ref{GeneralSol2}).
	
	In Fig.~\ref{real-case-deleteone} we can see plots of the potentials generated by this transformation (Fig.~\ref{subfig-1:real-case-deleteone}) and the seed solutions that were employed (Figs.~\ref{subfig-2:real-case-deleteone} and \ref{subfig-3:real-case-deleteone}).\\
	
	\begin{figure}[h]
		\centering
		\captionsetup[subfloat]{labelfont=bf}
		\begin{tabular}{cc}
			\adjustbox{valign=b}{
				\subfloat[\label{subfig-1:real-case-deleteone}]{%
					\includegraphics[scale=.9]{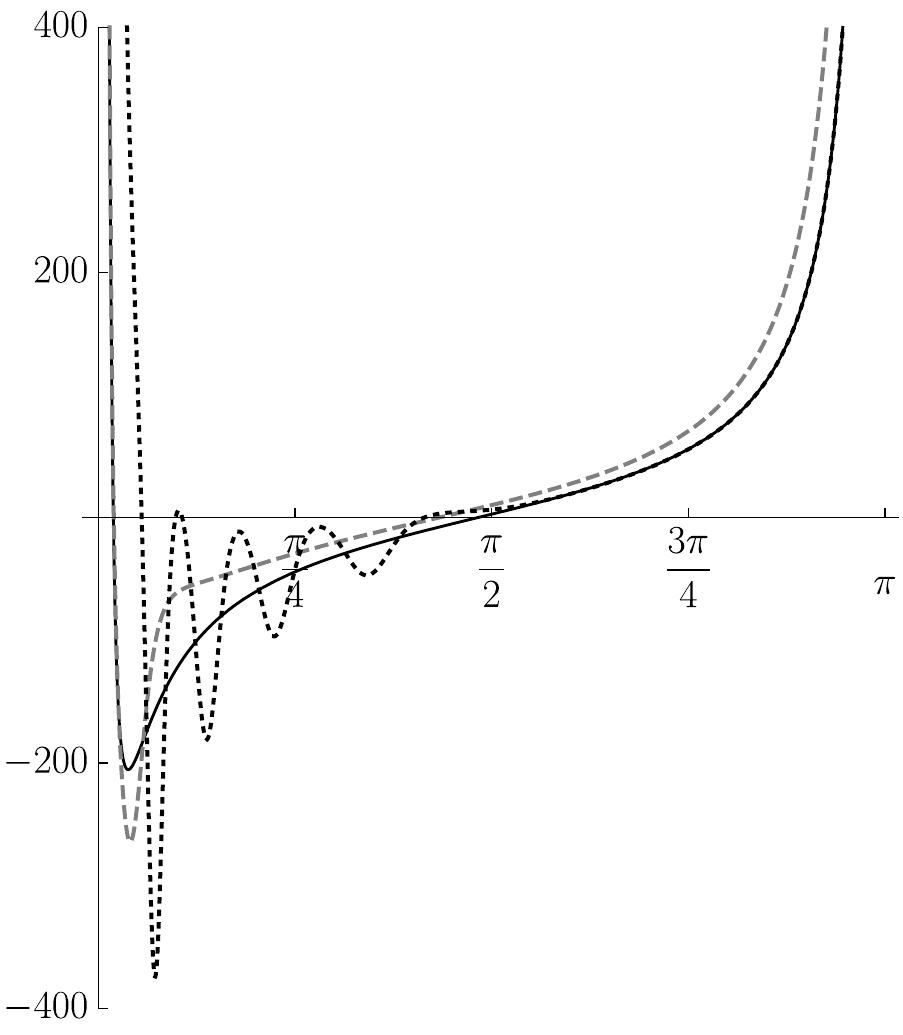}}}
			&      
			\adjustbox{valign=b}{\begin{tabular}{@{}c@{}}
					$\qquad$\subfloat[Seed solutions \mbox{$u_{01}(x)=\psi_{01}(x)$} (black curve) and \mbox{$u_{02}(x)=\psi_R (x)$} for \mbox{$\epsilon_2=-100$} (gray curve).\label{subfig-2:real-case-deleteone}]{%
						\includegraphics[scale=.65]{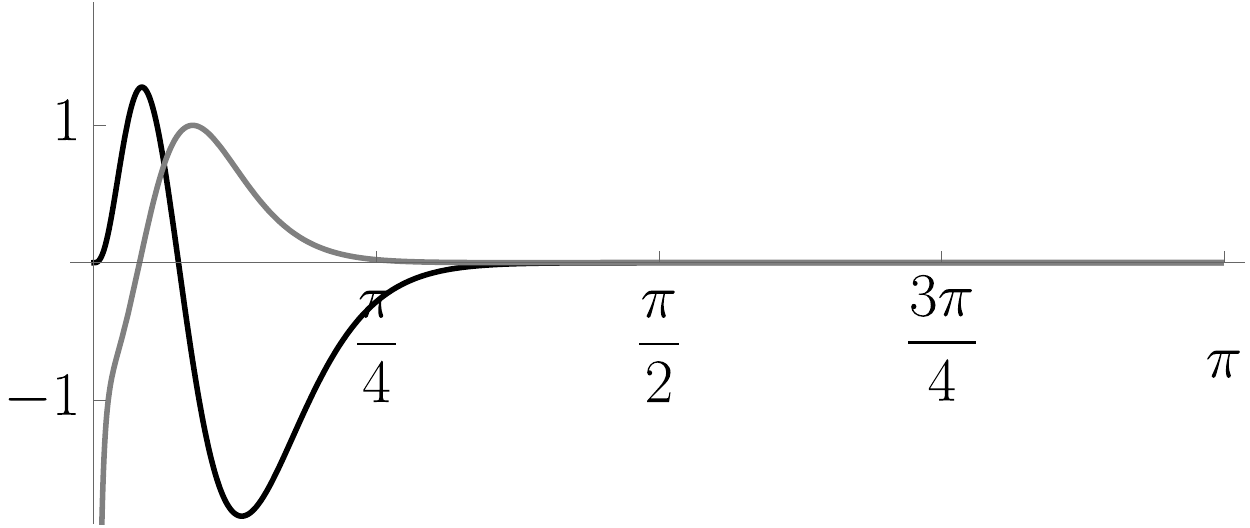}} \\
					$\qquad$\subfloat[Seed solutions \mbox{$u_{01}(x)=\psi_L (x)$} for \mbox{$\epsilon_1=0$} (black curve) and \mbox{$u_{02}(x)=\psi_{04}(x)$} (gray curve).\label{subfig-3:real-case-deleteone}]{%
						\includegraphics[scale=.65]{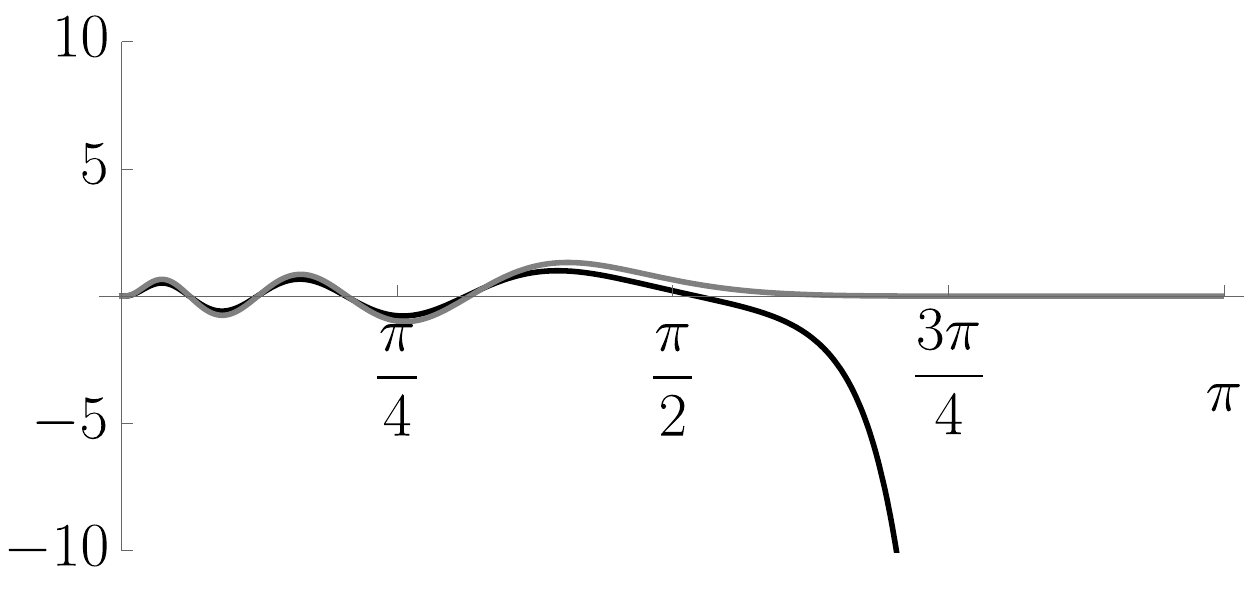}}
			\end{tabular}}
		\end{tabular}
		\caption{Deleting one energy level. \textbf{(a)} TRM potential for $a=2$, $b=50$ (black solid curve) and its second-order SUSY partners $V_2(x)$ for $\epsilon_1=E_1=-70.125$, $\epsilon_2=-100$ (gray dashed curve) and $\epsilon_1=0$, $\epsilon_2=E_4=-1.01$ (black dotted curve). Seed solutions used to plot the gray dashed curve \textbf{(b)} and the black dotted curve \textbf{(c)} of $V_2(x)$.\label{real-case-deleteone}}
	\end{figure}

	\item[(vi)] Isospectral transformations 	
	
	It is possible to generate as well Hamiltonians which are isospectral to the initial one. This is achieved for factorization energies which are outside the initial spectrum, in the gap $(E_j,E_{j+1})$, with the seed solutions being taken either as $\psi_{L}$ or $\psi_{R}$. 
	
	The new potential is given by Eq.~(\ref{V2gen}), since if one seed solution is $\psi_{L}$ and the other is $\psi_{R}$, the corresponding null and divergent behaviors cancel out at both ends of the $x$-domain, without changing the coefficient of the term $\csc^2(x)$. On the other hand, if both seed solutions are taken as $\psi_{L}$ ($\psi_{R}$), then the singularity induced on $V_2(x)$ in the neighborhood of $x=0$ is opposite to what happens around $x=\pi$, thus the coefficient of the term $\csc^2(x)$ changes in different ways at both ends.
	
	In Fig.~\ref{real-case-isospectral} we can see plots of potentials obtained by this transformation (Fig.~\ref{subfig-1:real-case-isospectral}). The potential in the gray dashed curve is generated using both seed solutions of type $\psi_{L}$ (Fig.~\ref{subfig-2:real-case-isospectral}) while for the potential in the black dotted curve one seed solution is of type $\psi_{L}$ and the other of type $\psi_{R}$ (Fig.~\ref{subfig-3:real-case-isospectral}). \\
	
	\begin{figure}
		\centering
		\captionsetup[subfloat]{labelfont=bf}
		\begin{tabular}{cc}
			\adjustbox{valign=b}{
				\subfloat[\label{subfig-1:real-case-isospectral}]{%
					\includegraphics[scale=.9]{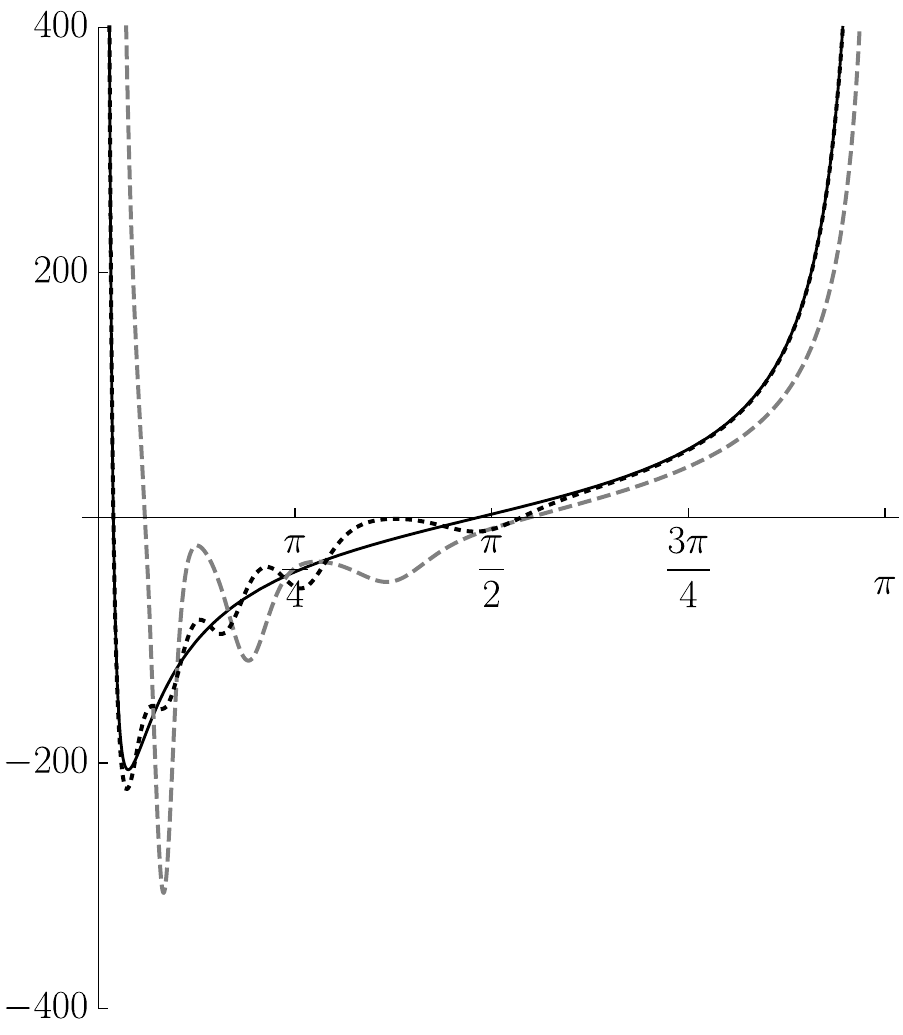}}}
			&      
			\adjustbox{valign=b}{\begin{tabular}{@{}c@{}}
					$\qquad$\subfloat[Seed solutions \mbox{$u_{01}(x)=\psi_L (x)$} for \mbox{$\epsilon_1=-40$} (black curve) and \mbox{$u_{02}(x)=\psi_L (x)$} for \mbox{$\epsilon_2=-60$} (gray curve).\label{subfig-2:real-case-isospectral}]{%
						\includegraphics[scale=.62]{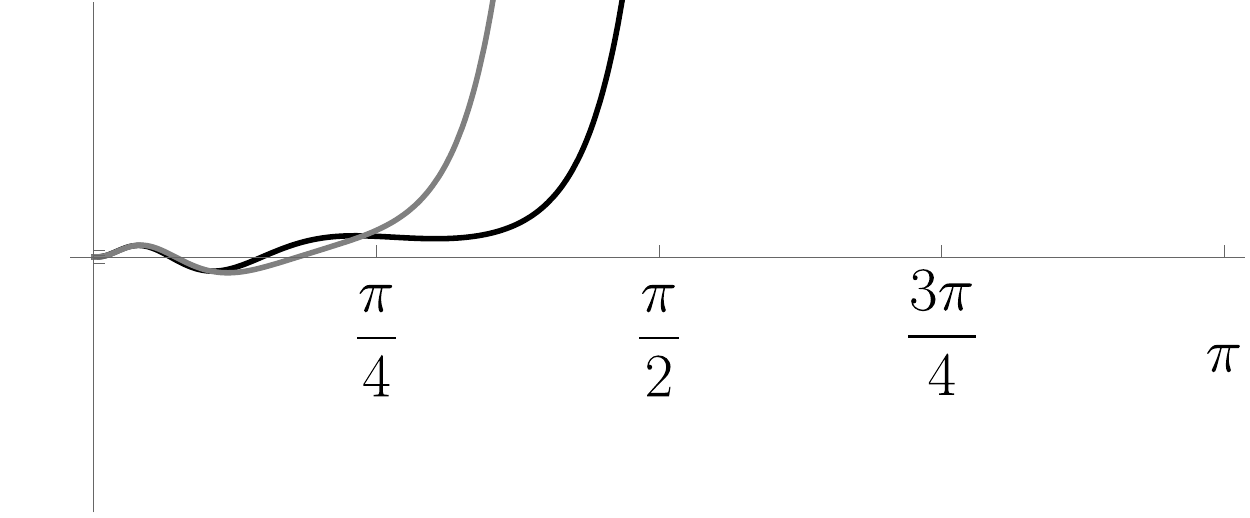}} \\
					$\qquad$\subfloat[Seed solutions \mbox{$u_{01}(x)=\psi_L (x)$} for \mbox{$\epsilon_1=-2$} (black curve) and \mbox{$u_{02}(x)=\psi_R (x)$} for \mbox{$\epsilon_2=-10$} (gray curve).\label{subfig-3:real-case-isospectral}]{%
						\includegraphics[scale=.62]{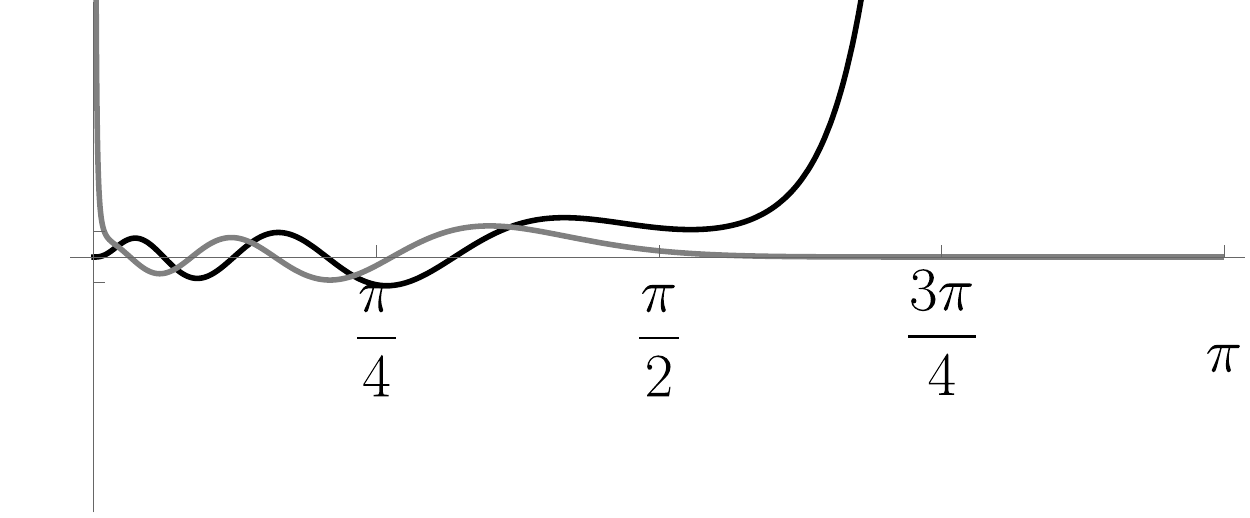}}
			\end{tabular}}
		\end{tabular}
		\caption{Isospectral transformations. \textbf{(a)} TRM potential for $a=2$, $b=50$ (black solid curve) and its second-order SUSY partners $V_2(x)$ for $\epsilon_1=-40$, $\epsilon_2=-60$ (gray dashed curve) and $\epsilon_1=-2$, $\epsilon_2=-10$ (black dotted curve). Seed solutions used to plot the gray dashed curve \textbf{(b)} and the black dotted curve \textbf{(c)} of $V_2(x)$.\label{real-case-isospectral}}
	\end{figure}

\end{itemize} 

\subsubsection{Complex case $(c<0)$}

The second-order SUSY transformation with two complex conjugate factorization energies supplies another way to generate isospectral Hamiltonians. The restrictions the seed solution must obey depend on the domain of definition of the initial potential, in our case $(0,\pi)$. Thus, in order to produce non-singular transformations in such a domain, the transformation function $u_{01}(x)$ associated to $\epsilon_1\in\setbox0=\hbox{\rm C}\hbox{\hbox
	to0pt{\kern0.4\wd0\vrule height0.9\ht0}\box0}$ must fulfill
\begin{equation}
\lim_{x \to 0}u_{01}(x)=0,\qquad \text{or}\qquad\lim_{x \to \pi}u_{01}(x)=0.
\label{complexcondition}
\end{equation}
If this requirement is compared with the behavior shown in Eqs.~(\ref{Limit0}) and (\ref{Limit20}), we realize that $\psi_{L}(x)$ and $\psi_{R}(x)$ already fulfill this condition, and the expressions (\ref{SolHyp}) and (\ref{SolHyp1}) for such solutions remain valid for complex $E$. Thus, the seed solutions $\psi_{L}(x)$ and $\psi_{R}(x)$ employed in this section are complex, and the parameters used to modify the shape of $V_2(x)$ are the real and imaginary parts of the factorization energy.

\begin{figure}[h]
	\centering
	\captionsetup[subfloat]{labelfont=bf}
	\begin{tabular}{cc}
		\adjustbox{valign=b}{
			\subfloat[\label{subfig-1:complex-case}]{%
				\includegraphics[scale=.85]{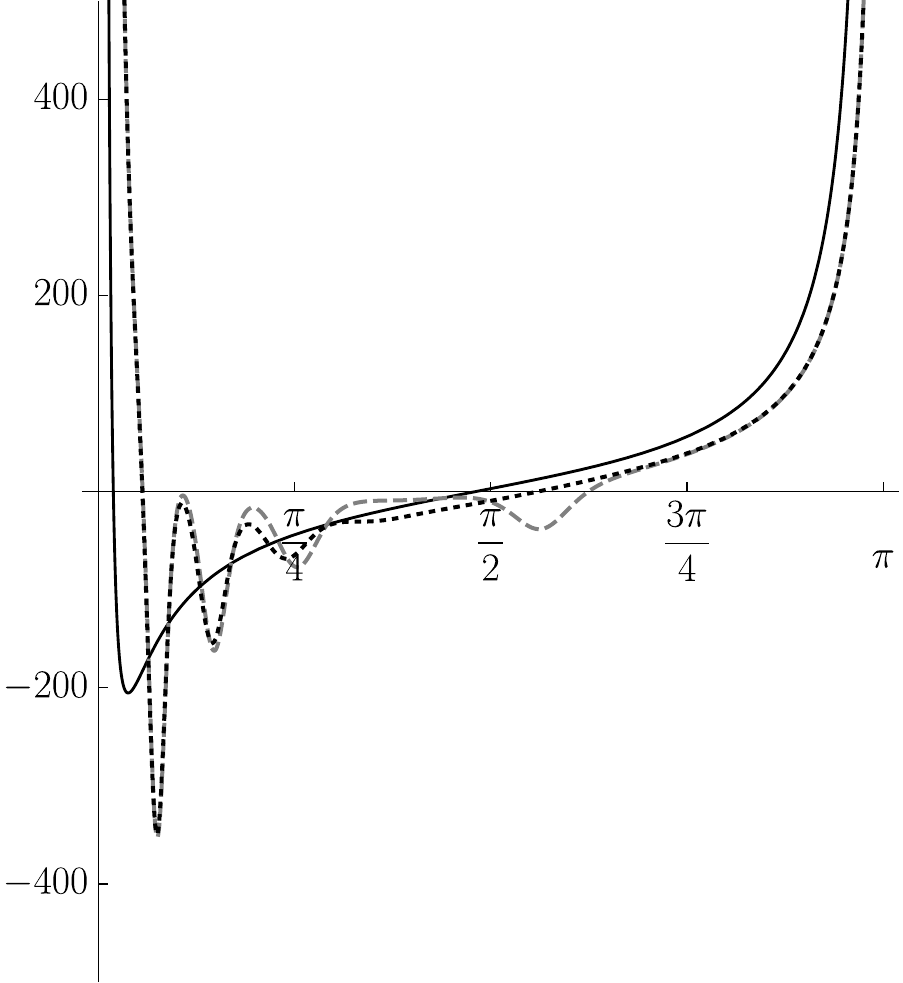}}}
		&      
		\adjustbox{valign=b}{\begin{tabular}{@{}c@{}}
				\subfloat[\label{subfig-2:complex-case}]{%
					\includegraphics[scale=.85]{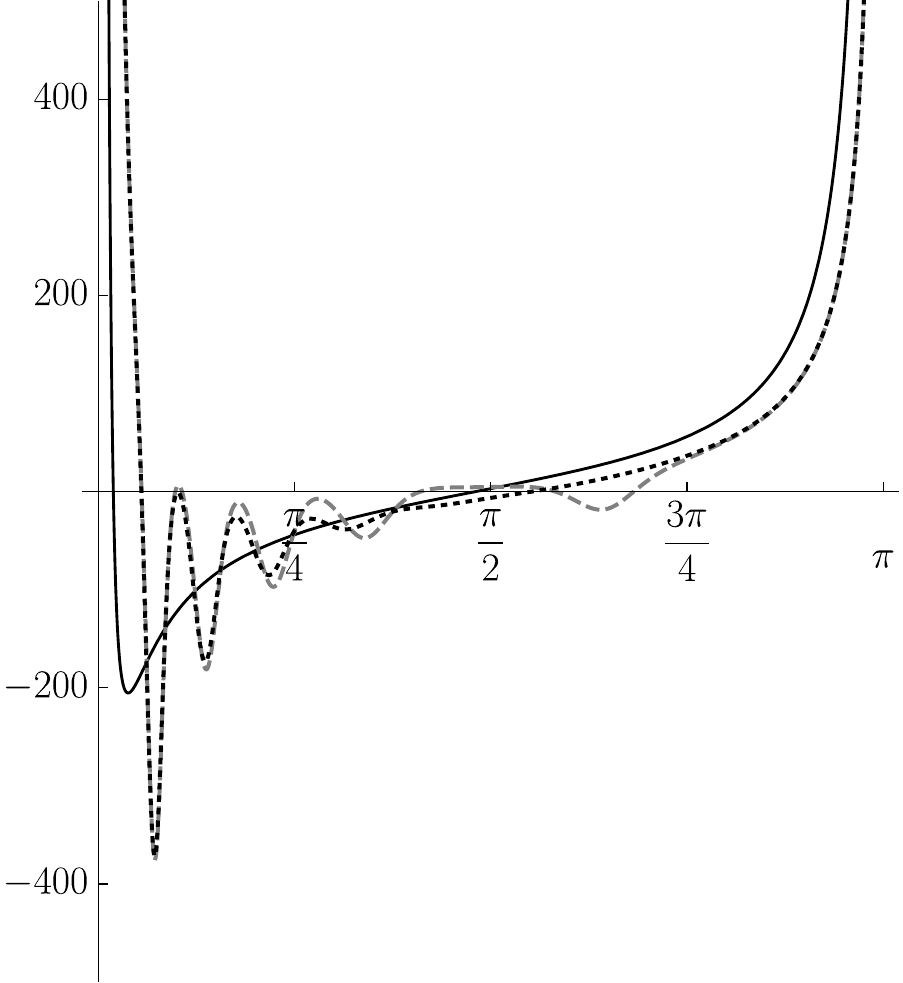}} 
		\end{tabular}}
	\end{tabular}
	\caption{Isospectral transformations. TRM potential for $a=2$, $b=50$ (black solid curve) and its second-order SUSY partners $V_2(x)$ (dashed and dotted curves) for the complex factorization energies: \textbf{(a)} $\epsilon_1=E_3+i=-16.72+i$ (gray dashed curve), and $\epsilon_1=E_3+20i=-16.72+20i$ (black dotted curve); \textbf{(b)} $\epsilon_1=i$ (gray dashed curve) and $\epsilon_1=20i$ (black dotted curve).\label{complex-case}}
\end{figure}

Some potentials generated through this transformation are plotted in Fig.~\ref{complex-case}. The initial TRM potential for $a=2$, $b=50$ (black solid curve) is drawn in all sub-figures, showing as well two potentials generated through second-order SUSY for a factorization energy $\epsilon_1$ with the same real part but different imaginary parts. Fig.~\ref{subfig-1:complex-case} shows the potentials generated for $\epsilon_1=E_3+i=-16.72+i$ (gray dashed curve) and $\epsilon_1=E_3+20i=-16.72+20i$ (black dotted curve), while Fig.~\ref{subfig-2:complex-case}  presents the potentials for  $\epsilon_1=i$ (gray dashed curve) and $\epsilon_1=20i$ (black dotted curve). As we can see, the maximum number of local minima the new potential shows seems to correspond with the number of nodes a real solution has in the gap within which the real part of $\epsilon_1$ falls. Let us stress also the way the imaginary part of $\epsilon_1$ smooths the oscillations in the SUSY partner potentials, as its absolute value grows up.

\subsubsection{Confluent case $(c=0)$}

The conditions the real seed solution $u_{01}$ must fulfill now are the same as those of the complex case previously discussed, for a factorization energy $\epsilon_1$ which now is real. Since for the TRM potential, by construction, the solutions $\psi_{L}(x)$ and $\psi_{R}(x)$ are real for $E\in \rm I\!R$, we can use them directly to implement the transformation. 
Let us note that in the confluent case it is possible to employ either physical or non-physical seed solutions, but in this work we will consider only bound states because the integrals for the non-physical seed solutions are more difficult to evaluate explicitly.

The confluent case allows to modify the initial spectrum in different ways, by changing the factorization energy and the constant $w_0$ in Eq.~(\ref{confluent}). Since here we will consider just factorization energies in $\text{Sp}(H_0)$, we can generate either isospectral Hamiltonians or delete one energy level of $\text{Sp}(H_0)$.

The potentials generated through the confluent algorithm depend on $W(u_{01},u_{02})$, which has an integral involved (see Eq.~(\ref{confluent})). For simplicity, let us take the lower limit of integration as the left end of the $x$-domain of the TRM potential, \textit{i.e.}, 
\begin{equation}
W(u_{01},u_{02})= w_0 + \int_{0}^x{\left[u_{01}(y)\right]^2\mbox{d}y}.
\label{confluentmod}
\end{equation} 
Thus, for $u_{01}(x)= \psi_{0j}(x)$ the $w_0$-values producing non-singular transformations must belong to the set $(-\infty,-1]\cup[0,\infty)$. In particular, 
for $w_0\in(-\infty,-1)\cup(0,\infty)$ the initial and final Hamiltonians are isospectral, but for $w_0$ tending to $0$ or $-1$ the energy level $E_j$ is erased from $\text{Sp}(H_2)$.

\begin{figure}[h]
	\centering
	\captionsetup[subfloat]{labelfont=bf}
	\begin{tabular}{cc}
		\adjustbox{valign=b}{
			\subfloat[\label{subfig-1:confluent-case}]{%
				\includegraphics[scale=.85]{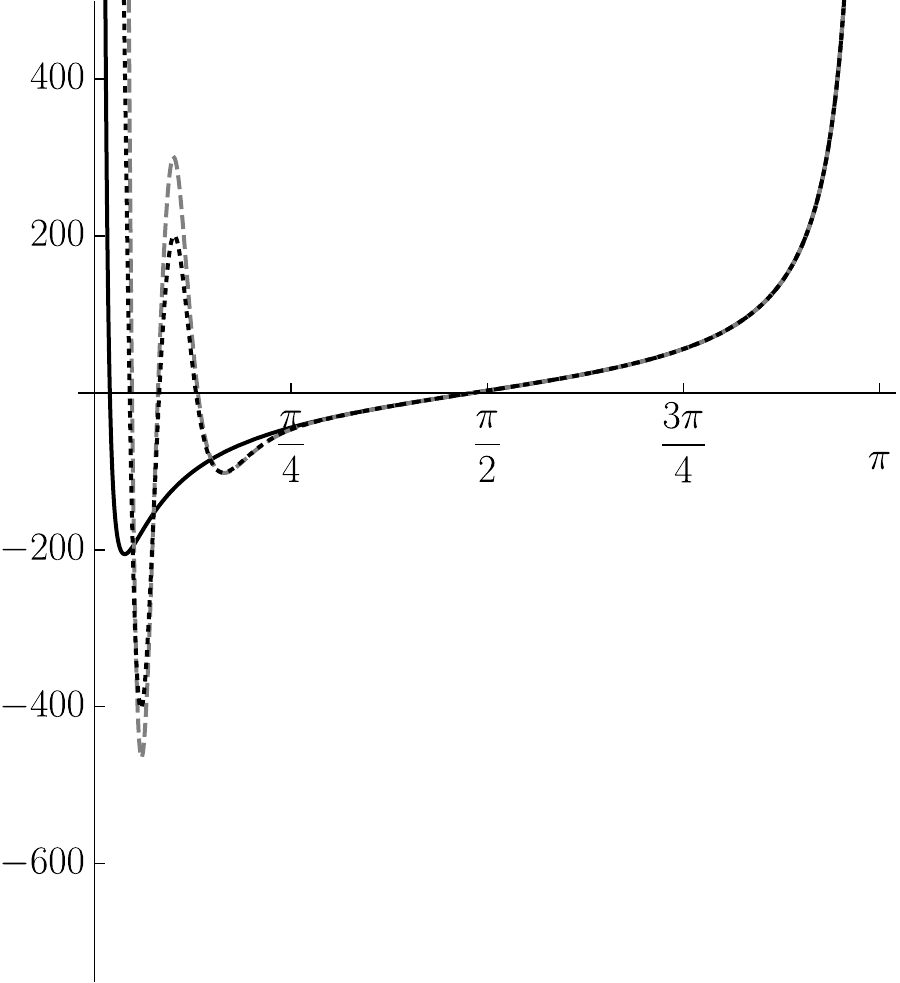}}}
		&      
		\adjustbox{valign=b}{\begin{tabular}{@{}c@{}}
				\subfloat[\label{subfig-2:confluent-case}]{%
					\includegraphics[scale=.85]{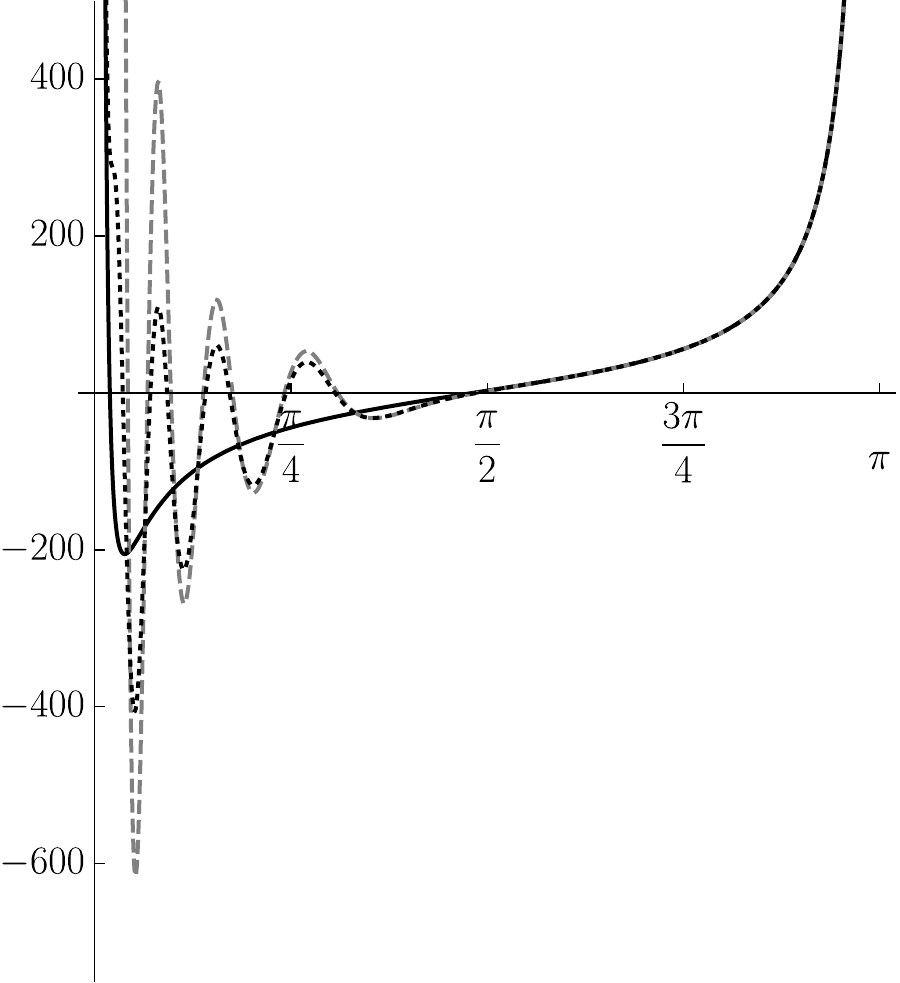}} 
		\end{tabular}}
	\end{tabular}
	\caption{Confluent second-order SUSY partner potentials $V_2(x)$ (dashed and dotted curves) of the TRM potential for $a=2$, $b=50$ (black solid curve) and: \textbf{(a)} $\epsilon_1=E_1=-70.125$, $w_0=0$ (gray dashed curve) and $w_0=0.05$ (black dotted curve); \textbf{(b)} $\epsilon_1=E_3=-16.71$, $w_0=0$ (gray dashed curve) and $w_0=0.05$ (black dotted curve).\label{confluent-case}}
\end{figure}

In Fig.~\ref{confluent-case} we can see plots of potentials generated through the confluent second-order SUSY applied to the TRM potential. The initial potential for $a=2$, $b=50$ (black solid curve) has been plotted in both sub-figures, and in each of them we are showing as well two potentials generated for a fixed $\epsilon_1$ and different $w_0$:  
Fig.~\ref{subfig-1:confluent-case} for $\epsilon_1=E_1=-70.125$, $w_0=0$ (gray dashed curve) and $w_0=0.05$ (black dotted curve); 
Fig.~\ref{subfig-2:confluent-case} for $\epsilon_1=E_3=-16.71$, $w_0=0$ (gray dashed curve) and $w_0=0.05$ (black dotted curve).

\section{Concluding remarks}
\label{Conclusions}

In this paper, the first and second-order SUSY transformations have been used to generate potentials with known spectra departing from {the TRM potentials. When we vary the parameters of the initial problem characterizing the SUSY transformation, it is possible to manipulate the final spectrum and to implement simple cases of spectral design. Some particular cases of first and second-order SUSY transformations applied to the TRM potential had been studied previously \cite{Dom11}. In this work we were able to go further, by expanding the cases which can be addressed and recovering the ones previously studied.
	
In order to generate the SUSY partners of the TRM potential, we have expressed the general solution to the initial stationary Schr\"odinger equation in terms of two appropriate linearly independent solutions $\psi_{R}(x)$ and $\psi_{L}(x)$, which vanish at the right and left ends of the potential domain, respectively. The elements of this basis set allow us to identify straightforwardly the terms that could generate divergences in the $x$-domain for the corresponding SUSY transformation. Moreover, the key feature of these solutions is that we can characterize in a precise way their asymptotic behavior at both ends of the $x$-domain. We have studied also the general solution to the stationary Schr\"odinger equation, as function of the coefficients of the linear combination and the associated energy parameter, thus getting relevant information about its global properties.
	
The general solution, expressed in terms of $\psi_{R}(x)$ and $\psi_{L}(x)$, allows us to identify simply the seed solutions with the required number of zeros for applying the non-singular transformations we are interested in. We have explored thoroughly the different types of first and second-order SUSY transformations that can be carried out, except when non-physical seed solutions of the TRM potential are involved in the confluent second-order SUSY transformation.
	
In order to finish, we want to point out that the potentials derived in this work are also of short range, as the TRM potentials. This means that they could be used as well in most of the physical situations described at the introduction of this paper, in particular as quark confinement models in QCD. Their main advantage, as compared with the TRM potentials, is the extra free parameters that are available, which could be used to fit better the physical quantities under study (see \textit{e.g.} \cite{dr97}). We hope that these ideas will be helpful for readers interested in SUSY QM and its physical applications.

\section*{Acknowledgments}
The authors acknowledge the support of Conacyt, project FORDECYT-PRONACES/61533/2020. Rosa Reyes also acknowledges the Conacyt scholarship  with number of CVU 280723.

\end{document}